\documentclass[prd,floatfix,amsmath,amssymb,nofootinbib,twocolumn]{revtex4-1}
\usepackage{graphicx,color,dcolumn,booktabs,bm}
\usepackage{longtable,comment,braket}
\usepackage{amsfonts,mathrsfs}
\usepackage{times}
\usepackage{overpic}
\usepackage{tipa}
\usepackage{indentfirst}
\usepackage{feynmf}   
\usepackage{slashed}  
\usepackage{cases}
\usepackage{epstopdf}
\usepackage{psfrag}
\usepackage{subfigure}
\usepackage{capt-of}
\usepackage{color}
\usepackage{multirow}
\usepackage{ulem}
\usepackage[colorlinks, citecolor=blue,anchorcolor=red,menucolor=red,linkcolor=blue,filecolor=red,runcolor=red,urlcolor=blue,frenchlinks=red]{hyperref}

\begin{document}
\title{$D^{(\ast)}N$ interaction and the structure of $\Sigma_c(2800)$ and $\Lambda_c(2940)$ in chiral effective field theory}
\author{Bo Wang$^{1,2}$}\email{bo-wang@pku.edu.cn}
\author{Lu Meng$^{2}$}\email{lmeng@pku.edu.cn}
\author{Shi-Lin Zhu$^{2,1}$}\email{zhusl@pku.edu.cn}
\affiliation{
$^1$Center of High Energy Physics, Peking University, Beijing 100871, China\\
$^2$ School of Physics and State Key Laboratory of Nuclear Physics
and Technology, Peking University, Beijing 100871, China}

\begin{abstract}
We study the $DN$ and $D^\ast N$ interactions to probe the inner
structure of $\Sigma_c(2800)$ and $\Lambda_c(2940)$ with the chiral
effective field theory to the next-to-leading order. We consider the
contact term, one-pion-exchange and two-pion-exchange contributions
to characterize the short-, long- and mid-range interactions of the
$D^{(\ast)}N$ systems. The low energy constants of the $D^{(\ast)}N$
systems are related to those of the $N\bar{N}$ interaction with
quark level Lagrangian that inspired by the resonance saturation
model. The $\Delta(1232)$ degree of freedom is also included in the
loop diagrams. The attractive potential in the $[DN]_{J=1/2}^{I=1}$
channel is too weak to form bound state, which indicates the
explanation of $\Sigma_c(2800)$ as the compact charmed baryon is
more reasonable. Meanwhile, the potentials of the isoscalar channels
are deep enough to yield the molecular states. We obtain the masses
of the $[DN]_{J=1/2}^{I=0}$, $[D^\ast N]_{J=1/2}^{I=0}$ and $[D^\ast
N]_{J=3/2}^{I=0}$ systems to be $2792.0$ MeV, $2943.6$ MeV and
$2938.4$ MeV, respectively. The $\Lambda_c(2940)$ is probably the
isoscalar $D^\ast N$ molecule considering its low mass puzzle.
Besides, the $\Lambda_c(2940)$ signal might contain the
spin-$\frac{1}{2}$ and spin-$\frac{3}{2}$ two structures, which can
qualitatively explain the significant decay ratio to $D^0p$ and
$\Sigma_c\pi$. We also study the $\bar{B}^{(\ast)}N$ systems and
predict the possible molecular states in the isoscalar channels. We
hope experimentalists could hunt for the open charmed molecular
pentaquarks in the $\Lambda_c^+\pi^+\pi^-$ final state.
\end{abstract}
\maketitle

\section{Introduction}\label{Introduction}
Hadron spectroscopy plays an important role in understanding the low
energy behaviors of QCD. Quark model is very successful in
describing the hadron spectra~\cite{Tanabashi:2018oca}. But it is
rather difficult to assign the near-threshold states, such as
$X(3872)$~\cite{Choi:2003ue} and $D_{s0}(2317)$~\cite{Aubert:2003fg}
to the quark model
predictions~\cite{Chen:2016qju,Guo:2017jvc,Liu:2019zoy,Lebed:2016hpi,Esposito:2016noz,Brambilla:2019esw}.
In the charmed baryon family, a state $\Lambda_c(2940)$ also falls
into the same situation as the $X(3872)$ and $D_{s0}(2317)$.

In 2007, the BaBar Collaboration observed a charmed baryon
$\Lambda_c(2940)$ in the $D^0p$ invariant mass
spectrum~\cite{Aubert:2006sp}, which is an isosinglet since no
signal is observed in the $D^+p$ final state. It was subsequently
confirmed by the Belle experiment in the decay mode
$\Lambda_c(2940)\to\Sigma_c\pi$~\cite{Abe:2006rz}. In 2017, the
$J^P$ quantum numbers of $\Lambda_c(2940)$ was constrained by the
LHCb measurement, and the most likely spin-parity assignment for
$\Lambda_c(2940)$ is $J^P=\frac{3}{2}^-$~\cite{Aaij:2017vbw} (The
mass and width of $\Lambda_c(2940)$ obtained by the BaBar, Belle and
LHCb experiments are shown in Table~\ref{LambdacInfo}).

\begin{table}[htbp]
\centering
\renewcommand{\arraystretch}{1.5}
\caption{The mass and width of $\Lambda_c(2940)$ in experiments (in
units of MeV).\label{LambdacInfo}} \setlength{\tabcolsep}{3.8mm}{
\begin{tabular}{ccc}
\hline
BaBar&$2939.8\pm1.3\pm1.0$&$17.5\pm5.2\pm5.9$\\
Belle&$2938.0\pm1.3^{+2.0}_{-4.0}$&$13^{+8+27}_{-5-7}$\\
LHCb&$2944.8^{+3.5}_{-2.5}\pm0.4^{+0.1}_{-4.6}$&$27.7^{+8.2}_{-6.0}\pm0.9^{+5.2}_{-10.4}$\\
\hline
\end{tabular}
}
\end{table}

Up to now, there are two different interpretations of the internal
structure of $\Lambda_c(2940)$. One is the ordinal charmed baryon,
and the other one is the $D^\ast N$ molecular state. However, it is
difficult to arrange $\Lambda_c(2940)$ to the $2P$ state in the
charmed baryon spectroscopy, since its mass is about $60$$-$$100$
MeV smaller than the calculations of the quark
models~\cite{Capstick:1986bm,Ebert:2011kk,Chen:2014nyo,Lu:2016ctt}.
Considering the $\Lambda_c(2940)$ lies about $6$ MeV below the
$D^{\ast0}p$ threshold, the molecular explanation was firstly
proposed in Ref.~\cite{He:2006is}, where the $\Lambda_c(2940)$ as
the $\frac{1}{2}^-$ molecular state is preferred by analyzing its
decay behaviors. In Ref.~\cite{He:2010zq}, He {\it et al}. studied
the $D^\ast N$ interaction with the one-boson-exchange model, and
their calculation supports the interpretation of the
$\Lambda_c(2940)$ as the $D^\ast N$ bound state with
$I(J^P)=0(\frac{1}{2}^+)$ or $0(\frac{3}{2}^-)$. In
Ref.~\cite{Ortega:2012cx}, Ortega {\it et al}. investigated the
$\Lambda_c(2940)$ as a $D^\ast N$ molecule in the constituent quark
model, and they obtain the binding solution in isoscalar
$J^P=\frac{3}{2}^-$ channel. In
Refs.~\cite{Dong:2009tg,Dong:2010xv}, the strong and radiative
decays of $\Lambda_c(2940)$ are calculated in the molecular picture.
A QCD sum rule study in Ref.~\cite{Zhang:2012jk} indicates the
$\Lambda_c(2940)$ is not a compact state. Some recent calculations
based on the chiral quark model also support the molecular
explanation for $\Lambda_c(2940)$~\cite{Zhao:2016zhf,Zhang:2019vqe}.
(see
Refs.~\cite{Klempt:2009pi,Crede:2013sze,Cheng:2015iom,Chen:2016spr,Kato:2018ijx}
for review and
Refs.~\cite{Huang:2016ygf,Wang:2015rda,Cheng:2015naa,Xie:2015zga,Romanets:2012hm,He:2011jp,Cheng:2006dk,Chen:2007xf,Zhong:2007gp,Chen:2015kpa}
for other related works).

Another charmed baryon related with the $DN$ threshold is
$\Sigma_c(2800)$, which is an isotriplet and firstly observed by the
Belle Collaboration in the $\Lambda_c\pi$ mass
spectrum~\cite{Mizuk:2004yu}. The  neutral state $\Sigma_c(2800)^0$
was possibly confirmed by the BaBar experiment~\cite{Aubert:2008ax},
but the measured mass from BaBar is about $50$ MeV larger. The $J^P$
of $\Sigma_c(2800)$ is still undetermined
yet~\cite{Tanabashi:2018oca}. Like the $\Lambda_c(2940)$,
$\Sigma_c(2800)$ is interpreted as the $P$-wave excitation of the
charmed baryon in the $\lambda$
mode~\cite{Ebert:2011kk,Cheng:2006dk,Chen:2016iyi,Garcilazo:2007eh,Wang:2017kfr},
and $DN$
molecule~\cite{Zhang:2012jk,Zhao:2016zhf,Zhang:2019vqe,Dong:2010gu},
respectively.

Investigating the $DN$ and $D^\ast N$ interaction is essential to
disentangle the puzzles of $\Sigma_c(2800)$ and $\Lambda_c(2940)$.
Besides, understanding the $D^{(\ast)}N$ interaction is also crucial
to probe the $D$-mesic
nuclei~\cite{Tsushima:1998ru,GarciaRecio:2010vt} and the properties
of the charmed mesons in the nuclear
matter~\cite{Hosaka:2016ypm,Krein:2017usp}. An alternative approach
based on the meson-exchange model~\cite{Machleidt:1987hj} has been
employed to construct the $DN$ and $\bar{D}N$ interaction by the
J\"ulich
group~\cite{Haidenbauer:2007jq,Haidenbauer:2008ff,Haidenbauer:2010ch}.

Instead of the boson-exchange model, the modern theory of nuclear
force is built upon the pioneer work of
Weinberg~\cite{Weinberg:1990rz,Weinberg:1991um} and largely
developed in the framework of effective field theory. The chiral
effective field theory was extensively exploited to study the $NN$
interaction with great
success~\cite{Bernard:1995dp,Epelbaum:2008ga,Machleidt:2011zz,Meissner:2015wva,Hammer:2019poc,Machleidt:2020vzm}.
The chiral effective field theory was also utilized to study the
systems with heavy flavors in
Refs.~\cite{Liu:2012vd,Xu:2017tsr,Wang:2018atz,Meng:2019ilv,Wang:2019ato,Meng:2019nzy,Wang:2019nvm},
which is a powerful tool in predicting the $BB^\ast$ and $B^\ast
B^\ast$ bound states~\cite{Wang:2018atz}, reproducing the newly
observed pentaquarks~\cite{Wang:2019ato}, extrapolating the
$\Sigma_c N$ potential from lattice QCD result to the physical pion
mass~\cite{Meng:2019nzy}, and so on. As a natural extension of the
$NN$ interaction, in this work, we use the chiral effective field
theory to study the $D^{(\ast)} N$ interaction up to the
next-to-leading order. We simultaneously consider the long-, mid-
and short-range interactions, and include the contribution of
$\Delta(1232)$ in the loops as an intermediate state. With the
chiral effective field theory, we calculate the $D^{(\ast)}N$
effective potentials and search for the possible bound states. The
numerical results can be compared with the experimental data of
$\Lambda_c(2940)$ and $\Sigma_c(2800)$ to see whether they are the
genuine charmed baryons or the molecular nature.

This paper is organized as follows. In Sec.~\ref{Lagrangians}, we give the Lagrangians and effective potentials of the $D^{(\ast)} N$ systems. In Sec.~\ref{NumericalResults}, we illustrate our numerical results and discussions. In Sec.~\ref{Summary}, we conclude with a short summary. In Appendix~\ref{app:QM}, we relate the low energy constants to those of the $N\bar{N}$ system with quark model.

\section{Lagrangians and effective potentials}\label{Lagrangians}
\subsection{Effective chiral Lagrangians}\label{chiralLagrangians}

We first show the leading order Lagrangian of the nucleon and pion
interaction under the heavy baryon reduction~\cite{Scherer:2002tk},
which reads
\begin{eqnarray}
\mathcal{L}_{\mathcal{N}\varphi}=\bar{\mathcal{N}}(iv\cdot
\mathcal{D}+2g_a\mathcal{S}\cdot u)\mathcal{N},
\end{eqnarray}
where $\mathcal{N}=(p,n)^T$ denotes the the large component of the
nucleon field under the nonrelativistic reduction. $v=(1,\bm{0})$ is
the 4-velocity of the nucleon and
$\mathcal{D}_\mu=\partial_\mu+\Gamma_\mu$. $g_a\simeq1.29$ is the
axial-vector coupling constant.
$\mathcal{S}^\mu=\frac{i}{2}\gamma_5\sigma^{\mu\nu}v_\nu$ stands for
the spin operator of the nucleon. $\Gamma_\mu$ and $u_\mu$ are the
chiral connection and axial-vector current, respectively. Their
expressions read
\begin{eqnarray}
\Gamma_\mu&\equiv&\frac{1}{2}\left[\xi^\dag,\partial_\mu
\xi\right]\equiv\tau^i\Gamma_\mu^i,
u_\mu\equiv\frac{i}{2}\left\{\xi^\dag,\partial_\mu
\xi\right\}\equiv\tau^i\omega_\mu^i,
\end{eqnarray}
where $\tau^i$ is the Pauli matrix,
\begin{eqnarray}
\xi^2=U=\exp\left(\frac{i\varphi}{f_\pi}\right),\qquad
\varphi=\left[ \begin{array}{cc}
\pi^0&\sqrt{2}\pi^+\\
\sqrt{2}\pi^-&-\pi^0
\end{array} \right],
\end{eqnarray}
and $f_\pi=92.4$ MeV is the pion decay constant.

Considering the importance of $\Delta(1232)$ in the $NN$
interaction~\cite{Machleidt:1987hj,Holinde:1977rh,Krebs:2007rh,Epelbaum:2007sq,Kaiser:1998wa},
we adopt the small scale expansion method~\cite{Hemmert:1997ye} to
explicitly include the $\Delta(1232)$ in the Lagrangians. The
Lagrangian that delineates the $\Delta$-$N$-$\pi$ coupling is given
as
\begin{eqnarray}
\mathcal{L}_{\Delta\varphi}&=&-\bar{\mathcal{T}}_i^\mu(iv\cdot\mathcal{D}^{ij}-\delta^{ij}\delta_a+2g_1\mathcal{S}\cdot u^{ij})g_{\mu\nu}\mathcal{T}_j^\nu,\\
\mathcal{L}_{\Delta\mathcal{N}\varphi}&=&2g_\delta(\bar{\mathcal{T}}_i^\mu
g_{\mu\alpha}\omega_i^\alpha\mathcal{N}+\bar{\mathcal{N}}\omega_i^{\alpha\dagger}g_{\alpha\mu}\mathcal{T}_i^\mu),
\end{eqnarray}
where $\delta_a=m_\Delta-m_N$. $g_1=\frac{9}{5}g_a$ is estimated
with the quark model~\cite{Hemmert:1997ye}. $g_\delta\simeq1.05$ is
the coupling constant for $\Delta N\pi$ vertex. $\mathcal{T}_i^\mu$
denotes the spin-$\frac{3}{2}$ and isospin-$\frac{3}{2}$ field
$\Delta(1232)$ after performing the nonrelativistic reduction. Its
matrix form reads
\begin{eqnarray}
\mathcal{T}_\mu^1&=&\frac{1}{\sqrt{2}}\left[\begin{array}{c}
\Delta^{++}-\frac{1}{\sqrt{3}}\Delta^0\\
\frac{1}{\sqrt{3}}\Delta^+-\Delta^-
\end{array} \right]_\mu,\nonumber\\
\mathcal{T}_\mu^2&=&\frac{i}{\sqrt{2}}\left[\begin{array}{c}
\Delta^{++}+\frac{1}{\sqrt{3}}\Delta^0\\
\frac{1}{\sqrt{3}}\Delta^++\Delta^-
\end{array} \right]_\mu,
\mathcal{T}_\mu^3=-\sqrt{\frac{2}{3}}\left[\begin{array}{c}
\Delta^{+}\\
\Delta^0
\end{array} \right]_\mu.
\end{eqnarray}

The leading order Lagrangian that depicts the interaction between
the charmed mesons and light Goldstones
reads~\cite{Wise:1992hn,Manohar:2000dt}
\begin{eqnarray}\label{Meson_Lag_SF}
\mathcal{L}_{\mathcal{H}\varphi}=i\langle \mathcal{H}v
\cdot\mathcal{D}\bar{\mathcal{H}}\rangle
-\frac{1}{8}\delta_b\langle\mathcal{H}\sigma^{\mu \nu}
\bar{\mathcal{H}}\sigma_{\mu \nu}\rangle +g\langle\mathcal{H}
\slashed{u}\gamma_{5}\bar{\mathcal{H}}\rangle,
\end{eqnarray}
where $\langle\cdots\rangle$ represents the trace in spinor space.
$\delta_b$ is defined as $\delta_b=m_{D^\ast}-m_{D}$. $g\simeq-0.59$
stands for the axial coupling, whose sign is determined with the
help of quark model. The $\mathcal{H}$ is the super-field for the
charmed mesons, which reads
\begin{eqnarray}
\mathcal{H}&=&\frac{1+\slashed{v}}{2}\left(P_\mu^\ast\gamma^\mu+iP\gamma_5\right),\nonumber\\
\bar{\mathcal{H}}&=&\gamma^0H^\dag\gamma^0=\left(P_\mu^{\ast\dag}\gamma^\mu+iP^\dag\gamma_5\right)\frac{1+\slashed{v}}{2}.
\end{eqnarray}
with $P=(D^0,D^+)^T$ and $P^\ast=(D^{\ast0},D^{\ast+})^T$,
respectively. We construct the leading order contact Lagrangian to
describe the short distance interaction between the nucleon and
charmed meson,
\begin{eqnarray}\label{Lag_NH}
\mathcal{L}_{\mathcal{NH}}&=&D_a\bar{\mathcal{N}}\mathcal{N}\langle\bar{\mathcal{H}}\mathcal{H}\rangle+D_b\bar{\mathcal{N}}\gamma_\mu\gamma_5\mathcal{N}\langle\bar{\mathcal{H}}\gamma^\mu\gamma_5\mathcal{H}\rangle\nonumber\\
&&+E_a\bar{\mathcal{N}}\tau_i\mathcal{N}\langle\bar{\mathcal{H}}\tau_i\mathcal{H}\rangle+E_b\bar{\mathcal{N}}\gamma_\mu\gamma_5\tau_i\mathcal{N}\langle\bar{\mathcal{H}}\gamma^\mu\gamma_5\tau_i\mathcal{H}\rangle,\nonumber\\
\end{eqnarray}
where $D_a$, $D_b$, $E_a$ and $E_b$ are four low energy constants
(LECs). $D_a$ and $D_b$ contribute to the central potential and
spin-spin interaction, respectively. $E_a$ and $E_b$ are related
with the isospin-isospin interaction and contribute to the central
and spin-spin interaction in spin space, respectively. With the
quark model, we fix their values with the $N\bar{N}$ interaction as
inputs, which is given in the Appendix~\ref{app:QM}.

\subsection{Expressions of the effective potentials}\label{EffectivePotentials}
\begin{figure}[hptb]
\begin{centering}
    \scalebox{1.0}{\includegraphics[width=0.85\linewidth]{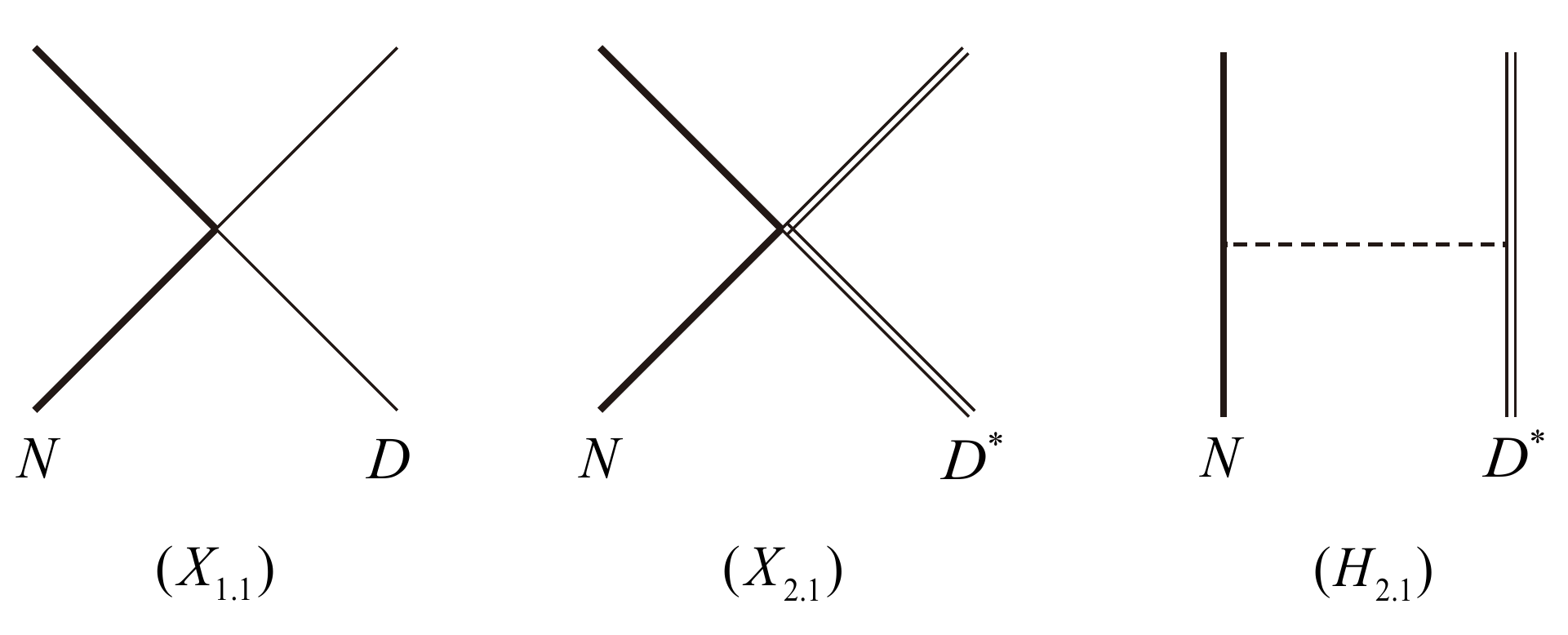}}
    \caption{The leading order Feynman diagrams that account for the $\mathcal{O}(\varepsilon^0)$ effective potentials of the $ND$ ($X_{1.1}$), and $ND^\ast$ ($X_{2.1},H_{2.1}$) systems. We use the thick, thin, double-thin and dashed lines to denote the $N$, $D$, $D^\ast$ and pion fields, respectively.\label{Tree_Level_Diagrams}}
\end{centering}
\end{figure}

In the framework of heavy hadron chiral perturbation theory, the
scattering amplitudes of the $D^{(\ast)}N$ systems can be expanded
order by order in powers of a small parameter
$\varepsilon=q/\Lambda_\chi$, where $q$ is either the momentum of
Goldstone bosons or the residual momentum of heavy hadrons, and
$\Lambda_\chi$ represents either the chiral breaking scale or the
mass of a heavy hadron\footnote{The mass splitting $\delta_b\sim m_\pi$,
so $\delta_b/\Lambda_\chi$ can be safely treated as the small parameter in chiral expansions.
However, the mass difference $\delta_a\sim 2 m_\pi\ll m_N$, so strictly speaking,
 the `small parameter' $\delta_a/\Lambda_\chi$ therefore has to be regarded as a
 {\it phenomenological} one. With the current accuracy, expanding to $(\delta_a/\Lambda_\chi)^2$
at the one-loop level can fulfill the convergence of the chiral expansions.}. The expansion is organized by the power
counting rule in Refs.~\cite{Weinberg:1990rz,Weinberg:1991um}. The
$\mathcal{O}(\varepsilon^0)$ Feynman diagrams for the $DN$ and
$D^\ast N$ systems are shown in Fig.~\ref{Tree_Level_Diagrams},
which contain the contact and one-pion-exchange diagrams. The
one-pion-exchange diagram for the $DN$ system vanishes since the
$DD\pi$ vertex is forbidden. The corresponding momentum-space
potentials of graphs in Fig.~\ref{Tree_Level_Diagrams} read
\begin{eqnarray}
\mathcal{V}_{DN}^{X_{1.1}}&=&D_a-4E_a(\mathbf{I}_1\cdot\mathbf{I}_2),
\end{eqnarray}
\begin{eqnarray}
\mathcal{V}_{D^\ast N}^{X_{2.1}}=D_a+D_b\boldsymbol{\sigma}\cdot\mathbf{T}-4(E_a+E_b\boldsymbol{\sigma}\cdot\mathbf{T})(\mathbf{I}_1\cdot\mathbf{I}_2),\label{Ct_DastN}\nonumber\\
\end{eqnarray}
\begin{eqnarray}
\mathcal{V}_{D^\ast
N}^{H_{2.1}}&=&(\mathbf{I}_1\cdot\mathbf{I}_2)\frac{gg_a}{f_\pi^2}\frac{(\bm{q}\cdot\boldsymbol{\sigma})(\bm{q}\cdot\mathbf{T})}{\bm{q}^2+m_\pi^2},
\end{eqnarray}
where $\mathbf{I}_1$ and $\mathbf{I}_2$ are the isospin operators of
$D$ and $N$, respectively. The operators $\boldsymbol{\sigma}$ and
$\mathbf{T}$ are related to the spin operators of the
spin-$1\over{2}$ baryon, spin-$1$ meson as
$\frac{1}{2}\boldsymbol{\sigma}$ and $-\mathbf{T}$, respectively
(see Ref.~\cite{Wang:2019ato} for details). The Breit approximation
\begin{eqnarray}
\mathcal{V}(\bm{q})=-\frac{\mathcal{M}(\bm{q})}{\sqrt{2\Pi_i m_i
2\Pi_f m_f}}
\end{eqnarray}
is used to relate the scattering amplitude $\mathcal{M}(\bm{q})$ to
the effective potential $\mathcal{V}(\bm{q})$ in momentum space
($m_i$ and $m_f$ are the masses of the initial and final states,
respectively).

\begin{figure*}[!hptb]
\begin{centering}
    \scalebox{1.0}{\includegraphics[width=0.85\linewidth]{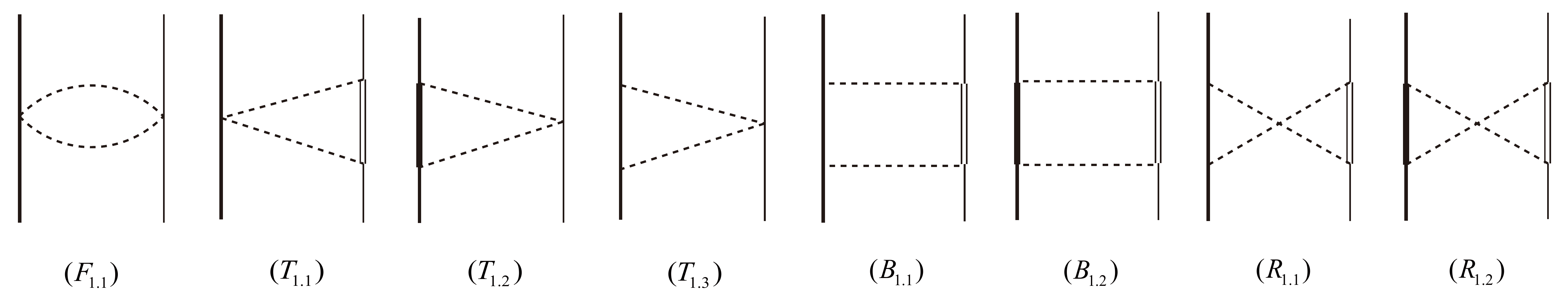}}
    \caption{The two-pion-exchange diagrams of the $DN$ system at $\mathcal{O}(\varepsilon^2)$. These diagrams are classified as the football diagram ($F_{1.1}$), triangle diagrams ($T_{1.i}$), box diagrams ($B_{1.i}$) and crossed box diagrams ($R_{1.i}$). We use the heavy-thick line to denote the $\Delta(1232)$ in the loops. Other notations are the same as those in Fig.~\ref{Tree_Level_Diagrams}.\label{TwoPion_Loop1}}
\end{centering}
\end{figure*}

The next-to-leading order two-pion-exchange diagrams for the $DN$
system are illustrated in Fig.~\ref{TwoPion_Loop1}. The effective
potentials from these graphs read
\begin{eqnarray}\label{eq14}
\mathcal{V}_{DN}^{F_{1.1}}&=&(\mathbf{I}_1\cdot\mathbf{I}_2)\frac{1}{f_\pi^4}
J_{22}^F(m_\pi,q),
\end{eqnarray}

\begin{eqnarray}
\mathcal{V}_{DN}^{T_{1.1}}&=&(\mathbf{I}_1\cdot\mathbf{I}_2)\frac{g^2}{f_\pi^4}\Big[(d-1) J_{34}^T\nonumber\\
&&-\bm{q}^2\left( J_{24}^T+
J_{33}^T\right)\Big](m_\pi,\mathcal{E}-\delta_b,q),
\end{eqnarray}
\begin{eqnarray}
\mathcal{V}_{DN}^{T_{1.2}}&=&(\mathbf{I}_1\cdot\mathbf{I}_2)\frac{4g_\delta^2}{3f_\pi^4}\Big[(2-d) J_{34}^T\nonumber\\
&&-\bm{q}^2\frac{2-d}{d-1}\left( J_{24}^T+
J_{33}^T\right)\Big](m_\pi,\mathcal{E}-\delta_a,q),
\end{eqnarray}
\begin{eqnarray}
\mathcal{V}_{DN}^{T_{1.3}}&=&(\mathbf{I}_1\cdot\mathbf{I}_2)\frac{g_a^2}{f_\pi^4}\Big[(d-1) J_{34}^T\nonumber\\
&&-\bm{q}^2\left( J_{24}^T+
J_{33}^T\right)\Big](m_\pi,\mathcal{E},q),
\end{eqnarray}
\begin{eqnarray}
\mathcal{V}_{DN}^{B_{1.1}}&=&\left[\frac{1}{8}-\frac{1}{3}\mathbf{I}_1\cdot\mathbf{I}_2\right]\frac{3g^2g_a^2}{2f_\pi^4}\Big[\bm{q}^4\left( J_{22}^B+2 J_{32}^B+ J_{43}^B\right)\nonumber\\
&&+(d^2-1) J_{41}^B
-2\bm{q}^2(d+1)\left( J_{31}^B+ J_{42}^B\right)\nonumber\\
&&-\bm{q}^2 J_{21}^B
\Big](m_\pi,\mathcal{E},\mathcal{E}-\delta_b,q),
\end{eqnarray}
\begin{eqnarray}
\mathcal{V}_{DN}^{B_{1.2}}&=&\left[\frac{1}{2}+\frac{2}{3}\mathbf{I}_1\cdot\mathbf{I}_2\right]\frac{g^2g_\delta^2}{f_\pi^4}\Big[\bm{q}^4\frac{d-2}{d-1}\left( J_{22}^B+2 J_{32}^B+ J_{43}^B\right)\nonumber\\
&&+(d^2-d-2) J_{41}^B
-2\bm{q}^2\frac{d^2-d-2}{d-1}\left( J_{31}^B+ J_{42}^B\right)\nonumber\\
&&-\bm{q}^2\frac{d-2}{d-1} J_{21}^B
\Big](m_\pi,\mathcal{E}-\delta_a,\mathcal{E}-\delta_b,q),
\end{eqnarray}
where the loop functions $J_{ij}^{F}$, $J_{ij}^{T}$, $J_{ij}^{B}$
and $J_{ij}^{R}$ are defined and given in
Refs.~\cite{Meng:2019ilv,Wang:2019ato,Wang:2018atz}. $d$ is the
dimension introduced in the dimensional regularization.
$\mathcal{E}=E_i-m_i~[i=N,D^{(\ast)}]$ represents the residual
energies of the $N$ and $D^{(\ast)}$. $\mathcal{E}$ is set to zero
in our calculations. The expressions of the crossed box diagrams
$(R_{1.i})$ can be obtained with the relation
\begin{eqnarray}
\mathcal{V}_{DN}^{R_{1.i}}&=&\mathcal{V}_{DN}^{B_{1.i}}\Big|_{
J_{ij}^B\to  J_{ij}^R,~\mathbf{I}_1\cdot\mathbf{I}_2\to
-\mathbf{I}_1\cdot\mathbf{I}_2}.
\end{eqnarray}
In order to generate the effective potential, one needs to subtract the two particle reducible contributions from the box diagrams. A detailed deduction that based on the principal value integral method is given in the Appendix B of Ref.~\cite{Wang:2019ato}.

\begin{figure*}[!hptb]
\begin{centering}
    \scalebox{1.0}{\includegraphics[width=0.85\linewidth]{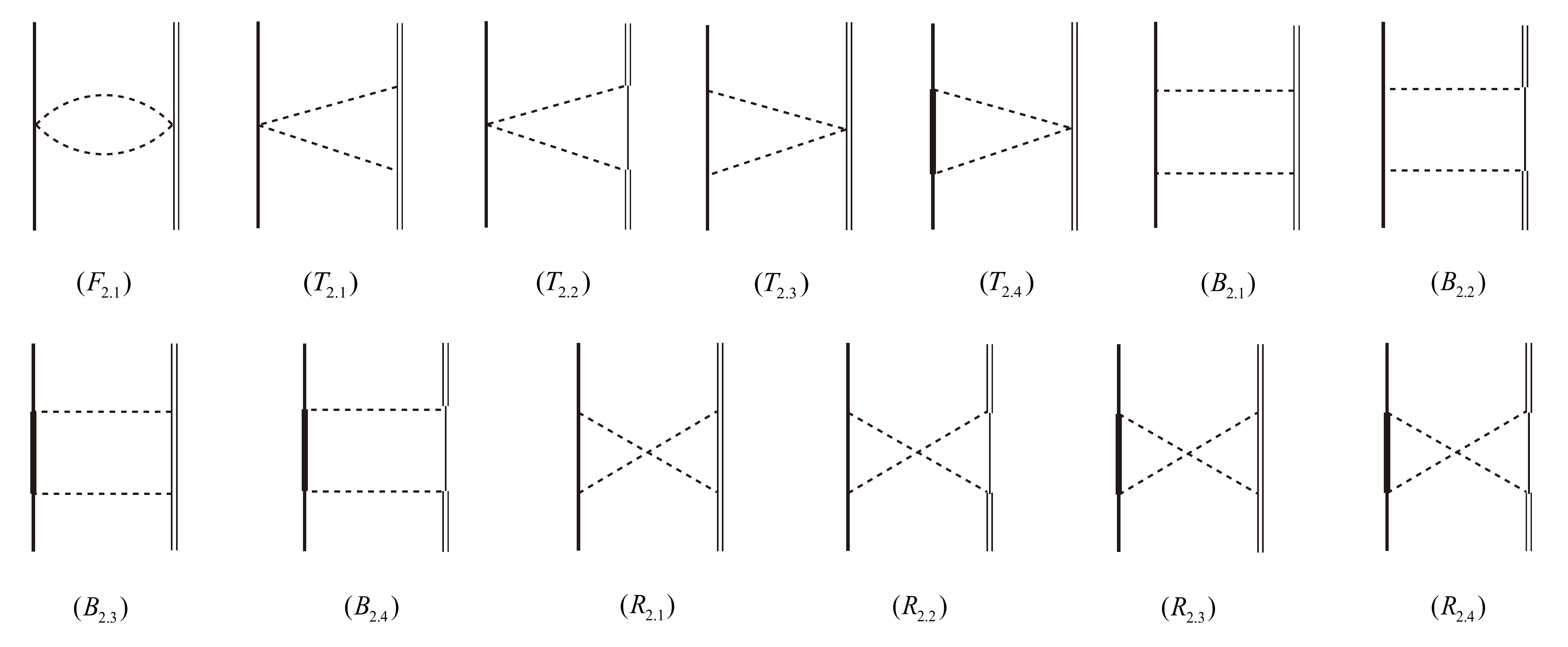}}
    \caption{The two-pion-exchange diagrams of the $D^\ast N$ system at $\mathcal{O}(\varepsilon^2)$. The notations are the same as those in Fig.~\ref{TwoPion_Loop1}.\label{TwoPion_Loop2}}
\end{centering}
\end{figure*}

The $\mathcal{O}(\varepsilon^2)$ two-pion-exchange diagrams for the
$D^\ast N$ system are shown in Fig.~\ref{TwoPion_Loop2}. Their
analytical expressions are written as
\begin{eqnarray}
\mathcal{V}_{D^\ast
N}^{F_{2.1}}&=&(\mathbf{I}_1\cdot\mathbf{I}_2)\frac{1}{f_\pi^4}
J_{22}^F(m_\pi,q),
\end{eqnarray}
\begin{widetext}
\begin{eqnarray}
\mathcal{V}_{D^\ast
N}^{T_{2.1}}&=&(\mathbf{I}_1\cdot\mathbf{I}_2)\frac{g^2}{f_\pi^4}\left[2
J_{34}^T-\bm{q}^2\frac{d-2}{d-1}\left( J_{24}^T+
J_{33}^T\right)\right](m_\pi,\mathcal{E},q),
\end{eqnarray}
\begin{eqnarray}
\mathcal{V}_{D^\ast
N}^{T_{2.2}}&=&(\mathbf{I}_1\cdot\mathbf{I}_2)\frac{g^2}{f_\pi^4}\left[
J_{34}^T-\frac{\bm{q}^2}{d-1}\left( J_{24}^T+
J_{33}^T\right)\right](m_\pi,\mathcal{E}+\delta_b,q),
\end{eqnarray}
\begin{eqnarray}
\mathcal{V}_{D^\ast
N}^{T_{2.3}}&=&(\mathbf{I}_1\cdot\mathbf{I}_2)\frac{g_a^2}{f_\pi^4}\bigg[(d-1)
J_{34}^T-\bm{q}^2\left( J_{24}^T+
J_{33}^T\right)\bigg](m_\pi,\mathcal{E},q),
\end{eqnarray}
\begin{eqnarray}
\mathcal{V}_{D^\ast
N}^{T_{2.4}}&=&(\mathbf{I}_1\cdot\mathbf{I}_2)\frac{4g_\delta^2}{3f_\pi^4}\bigg[(2-d)
J_{34}^T-\bm{q}^2\frac{2-d}{d-1}\left( J_{24}^T+
J_{33}^T\right)\bigg](m_\pi,\mathcal{E}-\delta_a,q),
\end{eqnarray}
\begin{eqnarray}
\mathcal{V}_{D^\ast N}^{B_{2.1}}&=&\left[\frac{1}{8}-\frac{1}{3}\mathbf{I}_1\cdot\mathbf{I}_2\right]\frac{3g^2g_a^2}{2f_\pi^4}\bigg[\frac{4d^2-10d+6}{d-1} J_{41}^B-\bm{q}^2\frac{d^2+3d-8}{d-1}\left( J_{31}^B+ J_{42}^B\right)\nonumber\\
&&-\bm{q}^2\frac{d-2+\boldsymbol{\sigma}\cdot\bm{T}}{d-1} J_{21}^B
+\bm{q}^4\frac{d-2}{d-1}\left( J_{22}^B+2 J_{32}^B+
J_{43}^B\right)\bigg](m_\pi,\mathcal{E},\mathcal{E},q),
\end{eqnarray}
\begin{eqnarray}
\mathcal{V}_{D^\ast
N}^{B_{2.2}}&=&\left[\frac{1}{8}-\frac{1}{3}\mathbf{I}_1\cdot\mathbf{I}_2\right]\frac{3g^2g_a^2}{2f_\pi^4}\bigg[-2\bm{q}^2\frac{d+1}{d-1}\left(
J_{31}^B+ J_{42}^B\right)
-\bm{q}^2\frac{1}{d-1}(1+\boldsymbol{\sigma}\cdot\bm{T}) J_{21}^B\nonumber\\
&&+(d+1) J_{41}^B+\bm{q}^4\frac{1}{d-1}\left( J_{22}^B+2 J_{32}^B+
J_{43}^B\right)\bigg](m_\pi,\mathcal{E},\mathcal{E}+\delta_b,q),
\end{eqnarray}
\begin{eqnarray}
\mathcal{V}_{D^\ast N}^{B_{2.3}}&=&\left[\frac{1}{2}+\frac{2}{3}\mathbf{I}_1\cdot\mathbf{I}_2\right]\frac{g^2g_\delta^2}{f_\pi^4}\bigg[-\bm{q}^2\frac{(d-2)^2-\boldsymbol{\sigma}\cdot\bm{T}}{(d-1)^2} J_{21}^B-\bm{q}^2\frac{(d-2)(d^2+3d-8)}{(d-1)^2}\left( J_{31}^B+ J_{42}^B\right)\nonumber\\
&&+\frac{2(d^2-2d+2)}{d-1}
J_{41}^B+\bm{q}^4\frac{(d-2)^2}{(d-1)^2}\left( J_{22}^B+2 J_{32}^B+
J_{43}^B\right)\bigg](m_\pi,\mathcal{E}-\delta_a,\mathcal{E},q),
\end{eqnarray}
\begin{eqnarray}\label{eq29}
\mathcal{V}_{D^\ast N}^{B_{2.4}}&=&\left[\frac{1}{2}+\frac{2}{3}\mathbf{I}_1\cdot\mathbf{I}_2\right]\frac{g^2g_\delta^2}{f_\pi^4}\frac{1}{d-1}\bigg[-2\bm{q}^2\frac{(d+1)(d-2)}{d-1}\left( J_{31}^B+ J_{42}^B\right)-\bm{q}^2\frac{d-2-\boldsymbol{\sigma}\cdot\bm{T}}{d-1} J_{21}^B\nonumber\\
&&+\bm{q}^4\frac{d-2}{d-1}\left( J_{22}^B+2 J_{32}^B+
J_{43}^B\right)+(d^2-d-2)
J_{41}^B\bigg](m_\pi,\mathcal{E}-\delta_a,\mathcal{E}+\delta_b,q).
\end{eqnarray}
\end{widetext}
The expressions of the diagrams $(R_{2.i})$ can be obtained with
\begin{eqnarray}
\mathcal{V}_{D^\ast N}^{R_{2.i}}=\mathcal{V}_{D^\ast
N}^{B_{2.i}}\Big|_{J_x^B\to
J_x^R,~\mathbf{I}_1\cdot\mathbf{I}_2\to-\mathbf{I}_1\cdot\mathbf{I}_2,~\boldsymbol{\sigma}\cdot\bm{T}\to-\boldsymbol{\sigma}\cdot\bm{T}}.
\end{eqnarray}
In above equations, the spin operator and transferred momentum are defined in $d$ dimensions, such as $\bm{S}^2=(d-1)/4$~\cite{Scherer:2002tk} and $q_iq_j=1/(d-1)\bm{q}^2\delta_{ij}$ for $S$-wave. The relation between the $\bm T$ operator and the polarization vector of $D^\ast$ meson can be found in the Appendix C of Ref.~\cite{Wang:2019ato}.

Notably, because of $\delta_b>m_\pi$, some diagrams in Fig.~\ref{TwoPion_Loop2}, such as $(T_{2.2})$ and $(B_{2.2})$, would also generate imaginary parts, which contribute to the width of the corresponding bound state. If solving the Lippmann-Schwinger equation, the imaginary part would shift the pole position in the Riemann sheet. In this work, we solve the Schr\"odinger equation to only focus on the binding energies, thus the imaginary parts are ignored in our calculations.

At the next-to-leading order, besides the two-pion-exchange potentials illustrated above, there also have the one-loop corrections to the one-pion-exchange and contact terms. These corrections can be replaced by using the physical values of the couplings, decay constant and masses of pion, $N$ and $D$ mesons, etc.. In addition, the subleading contact Lagrangians are also necessary. On the one hand, they are responsible for the renormalizations of two-pion-exchange loop diagrams. Because these diagrams are usually ultraviolet divergent in $d=4$ dimensions, we need the unrenormalized $\mathcal{O}(\varepsilon^2)$ LECs to absorb the divergent parts of the loop functions in Eqs.~\eqref{eq14}-\eqref{eq29}. On the other hand, these $\mathcal{O}(\varepsilon^2)$ Lagrangians may generate contact interactions with two powers of momenta in the following form~\cite{Machleidt:2011zz},
\begin{eqnarray}
\mathscr{V}&=&C_1\bm{q}^2+C_2\bm{k}^2+(C_3\bm{q}^2+C_4\bm{k}^2)\boldsymbol{\sigma}\cdot\bm{T}\nonumber\\
&&+C_5[i\bm{S}\cdot(\bm q\times \bm k)]+C_6(\boldsymbol{\sigma}\cdot\bm q)(\bm T\cdot \bm q)\nonumber\\
&&+C_7(\boldsymbol{\sigma}\cdot\bm k)(\bm T\cdot \bm k),
\end{eqnarray}
where $\bm k=(\bm{p}_N+\bm{p}_D)/2$ is the average momentum and $\bm S=(\bm{S}_N+\bm{S}_{D^\ast})/2$ is the total spin (Note that for $DN$ system, only the $C_1$, $C_2$ and $C_5$ terms survive). Unlike the $NN$ case, these $\mathcal{O}(\varepsilon^2)$ LECs cannot be fitted at present since the $D^{(\ast)}N$ scattering data is still unavailable. So we ignore their contributions and only consider the leading order contact terms. Once the $D^{(\ast)}N$ scattering phase shifts in lattice QCD simulations are available, it would be an intriguing topic to restudy this system with considering higher order contributions.

\section{Numerical results and discussions}\label{NumericalResults}
With the momentum-space potentials $\mathcal{V}(\bm{q})$ obtained in
Sec.~\ref{EffectivePotentials}, we make the following Fourier
transformation to get the effective potential $V(r)$ in the
coordinate space,
\begin{eqnarray}\label{Four_Tansf}
V(r)=\int\frac{d^3\bm{q}}{(2\pi)^3}e^{i\bm{q\cdot r}}\mathcal{V}(\bm
q)\mathcal{F}(\bm q).
\end{eqnarray}
We need to introduce a regulator $\mathcal{F}(\bm q)$ to suppress
the high momentum contribution. We choose the Gauss form
$\mathcal{F}(\bm{q})=\exp(-\bm{q}^{2n}/\Lambda^{2n})$ as used in the
$NN$ and $N\bar{N}$ systems~\cite{Epelbaum:2004fk,Epelbaum:2003xx}.
The Taylor expansion of the regulator function gives $\mathcal{F}(\bm{q})=1-\bm{q}^{2n}/\Lambda^{2n}+\dots$.
The power $n$ is chosen to be sufficiently large in order that
the cutoff induced contributions $\mathcal{V}(\bm q)\mathcal{O}(\bm{q}^{2n}/\Lambda^{2n})$ are beyond the chiral order one is working at.
In our calculation, we only consider the leading and subleading contributions, thus $n=1$ is already enough in our case. However,
we use the $N\bar{N}$ LECs fitted in Ref.~\cite{Kang:2013uia} to estimate the LECs of $D^{(\ast)}N$ systems, where the $n=3$ is adopted in Ref.~\cite{Kang:2013uia},
so we also choose $n=3$ for consistency. The power $n=3$ and cutoff $\Lambda\simeq0.5\pm0.1$ GeV are always
adopted to fit the experimental data and make predictions~\cite{Kang:2013uia,Epelbaum:2014efa,Epelbaum:2014efa,Entem:2003ft,Dai:2017ont}.

\subsection{Numerical results}

In order to get the numerical results, we also need to know the
values of the four LECs in Eq.~\eqref{Lag_NH}. Generally, these LECs
should be determined by fitting the $D^{(\ast)}N$ scattering data in
experiments or in lattice QCD simulations. However, the data in this
area are scarce, thus we have to resort to other alternative ways.
As proposed in Refs.~\cite{Meng:2019nzy,Wang:2019nvm}, we estimate
the LECs by constructing the contact Lagrangian at the quark level,
and then extract the couplings from the $N\bar{N}$ interaction,
which is demonstrated in Appendix~\ref{app:QM}.
\begin{figure*}
\begin{center}
\begin{minipage}[t]{0.32\linewidth}
\centering
\includegraphics[width=\columnwidth]{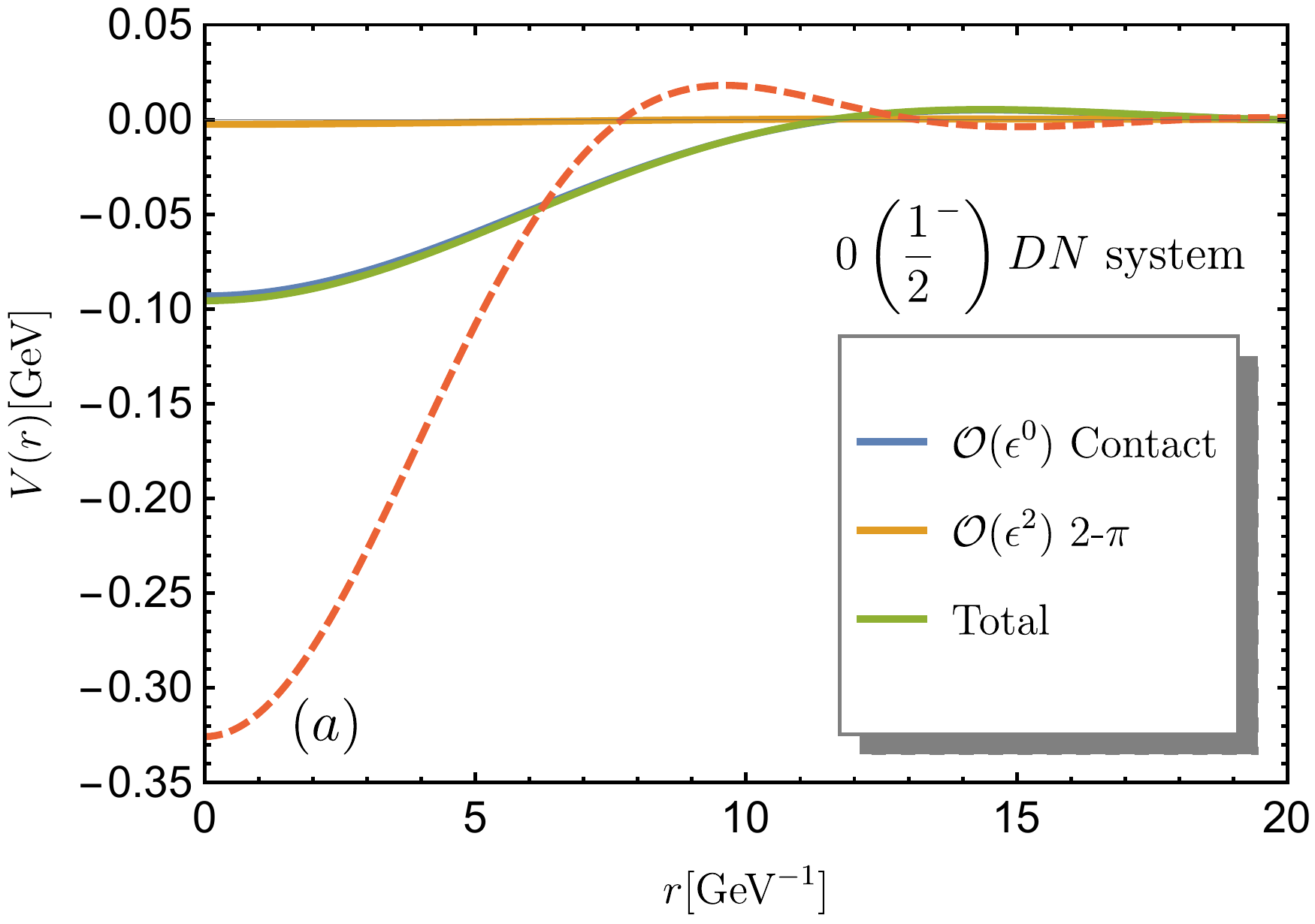}
\end{minipage}%
\begin{minipage}[t]{0.32\linewidth}
\centering
\includegraphics[width=\columnwidth]{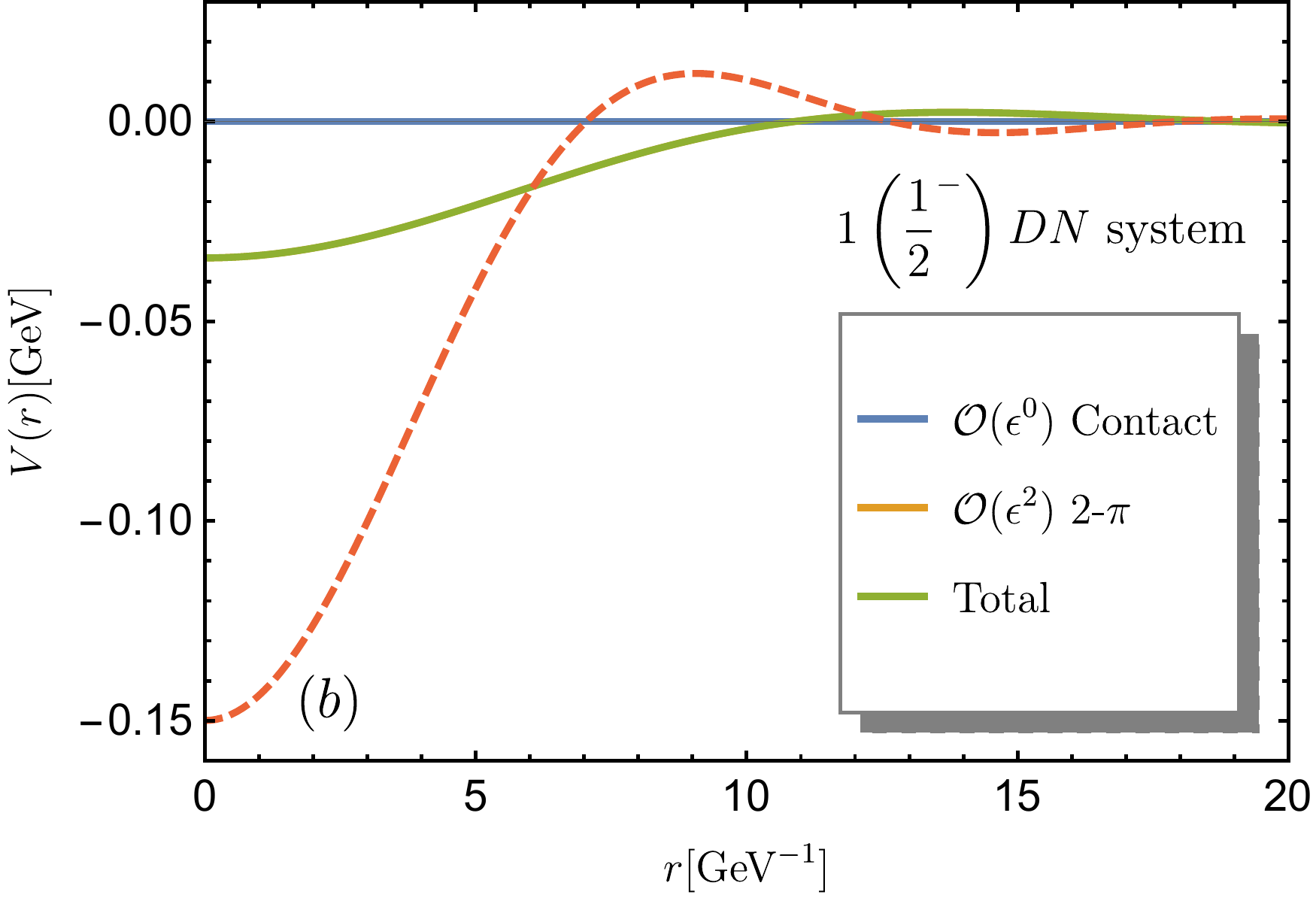}
\end{minipage}
\begin{minipage}[t]{0.32\linewidth}
\centering
\includegraphics[width=\columnwidth]{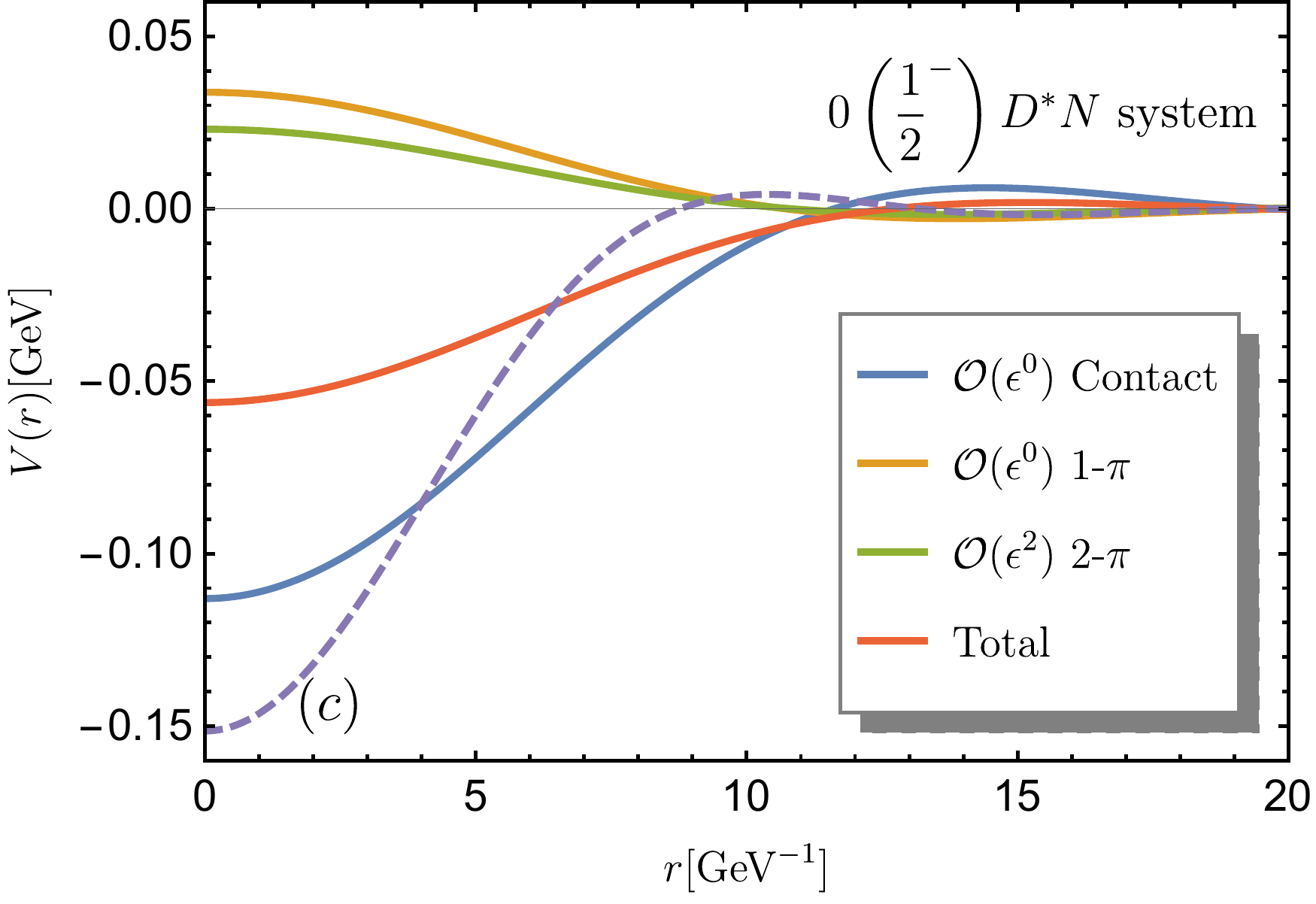}
\end{minipage}%
\\
\begin{minipage}[t]{0.32\linewidth}
\centering
\includegraphics[width=\columnwidth]{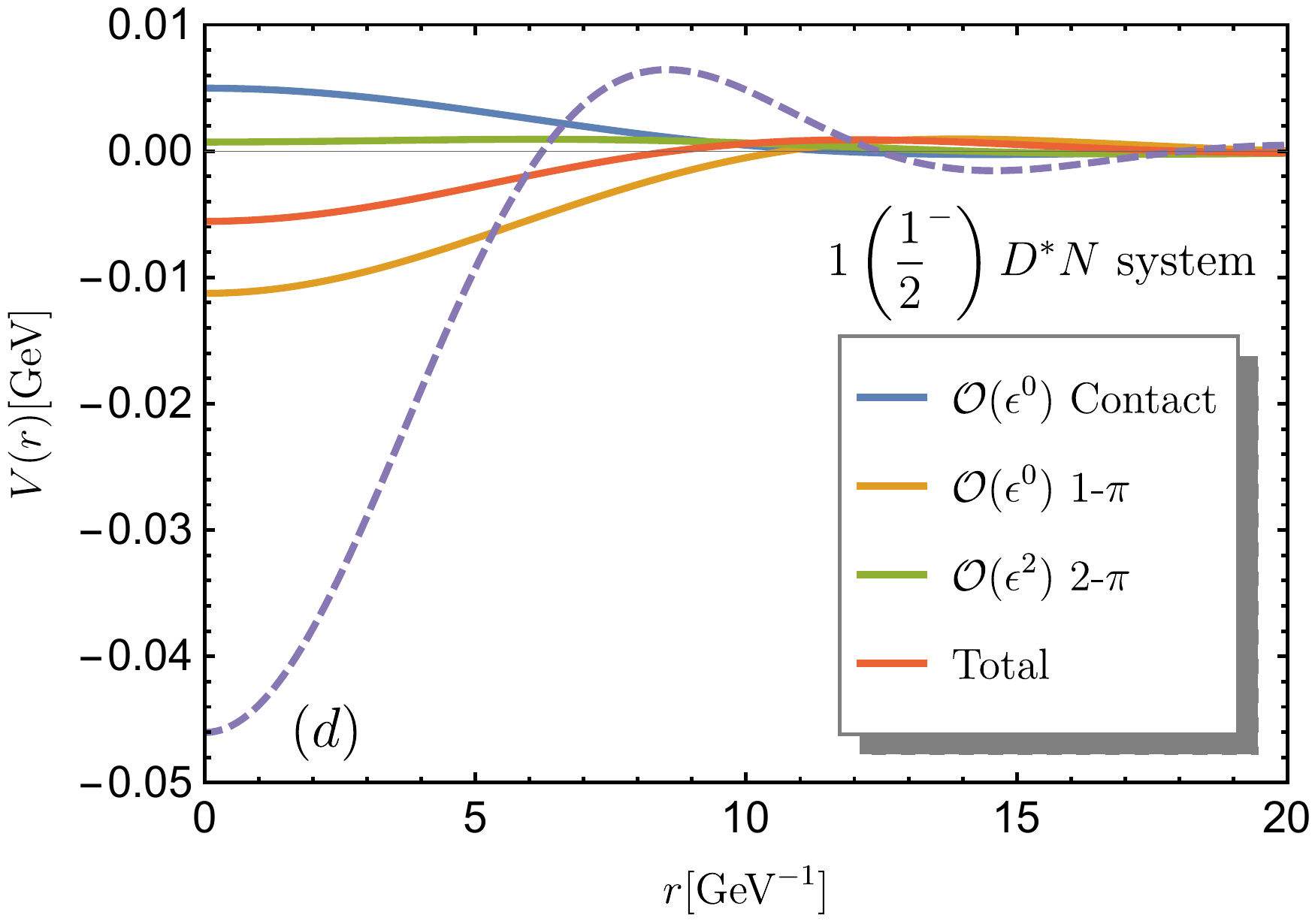}
\end{minipage}
\begin{minipage}[t]{0.32\linewidth}
\centering
\includegraphics[width=\columnwidth]{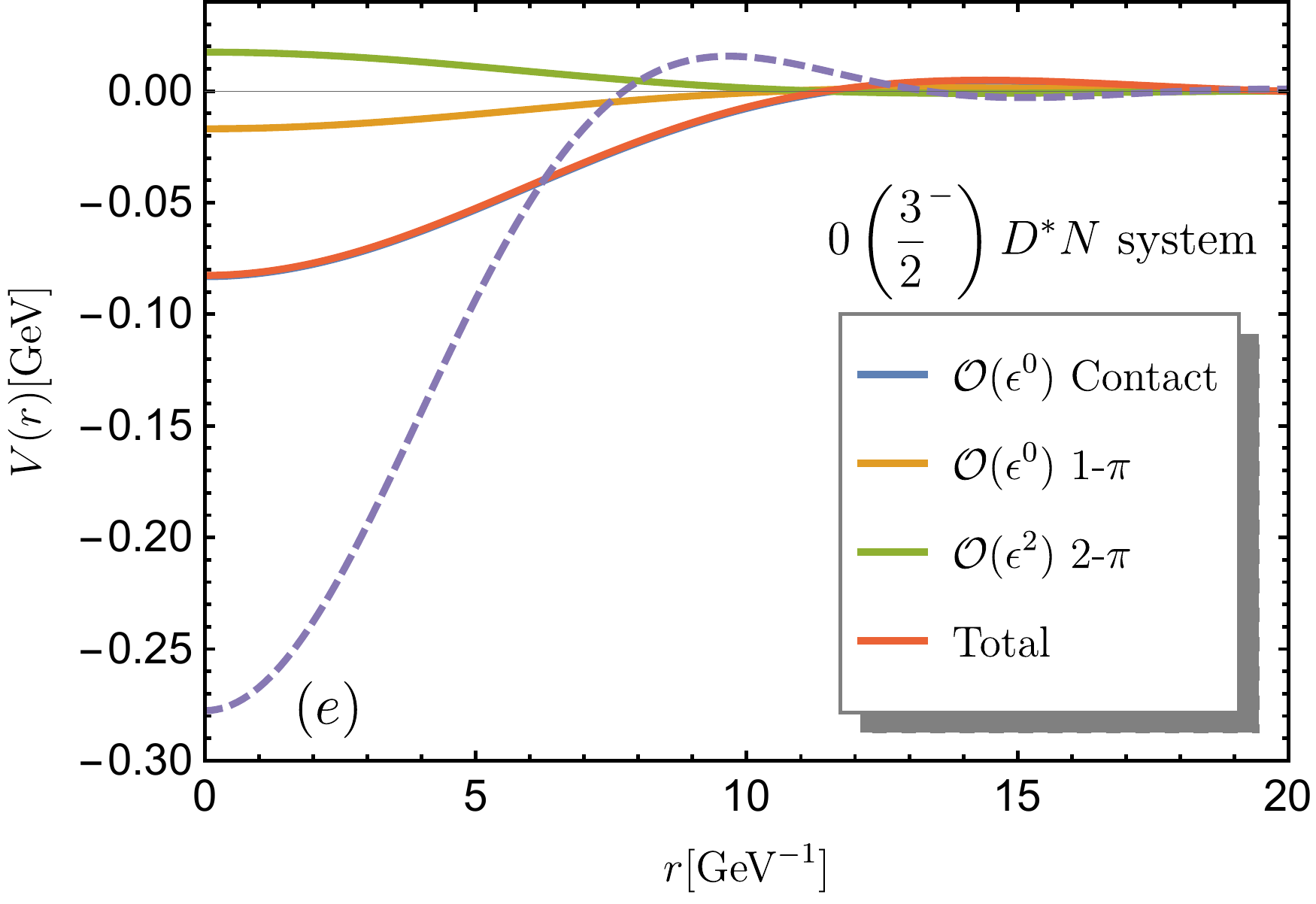}
\end{minipage}
\begin{minipage}[t]{0.32\linewidth}
\centering
\includegraphics[width=\columnwidth]{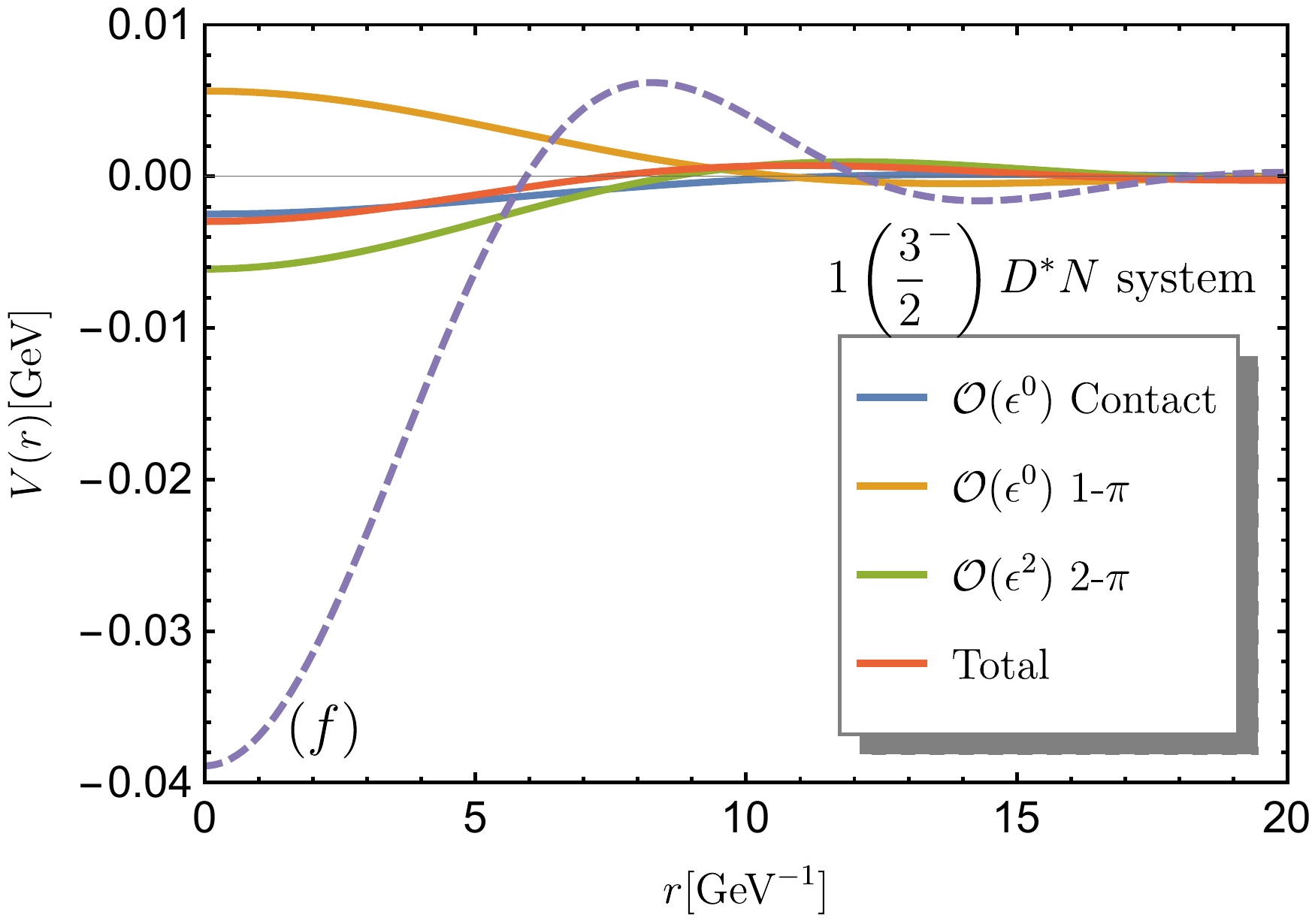}
\end{minipage}
\caption{The effective potentials of the $D^{(\ast)}N$ systems.
Their $I(J^P)$ are marked in each subfigure. The potentials with the solid lines are
obtained with the cutoff parameter $\Lambda=0.4$ GeV and the LECs
estimated in Appendix~\ref{app:QM}. The corresponding total potentials with the dashed lines are
obtained with the cutoff parameter $\Lambda=0.6$ GeV.\label{Potential2}}
\end{center}
\end{figure*}

We show the effective potentials of each possible $I(J^P)$
configurations in Fig.~\ref{Potential2}. In the following, we
analyze the behaviors of effective potentials for each system.

{\it $DN$ system}: The result in Fig.~\ref{Potential2}($a$) shows
that the $\mathcal{O}(\varepsilon^0)$ contact and
$\mathcal{O}(\varepsilon^2)$ two-pion-exchange potentials of the
$[DN]_{J=1/2}^{I=0}$ system are both attractive. But the attraction
of two-pion-exchange potential is rather weak. The attractive
potential is dominantly provided by the contact interaction. We find
a bound state in this channel. The binding energy and mass of this
state are predicted, respectively,
\begin{eqnarray}
\Delta E_{[D N]_{J=1/2}^{I=0}}&\simeq&-11.1~\textrm{MeV},\nonumber\\
M_{[D N]_{J=1/2}^{I=0}}&\simeq&2792.0~\textrm{MeV}.
\end{eqnarray}

For the $[DN]_{J=1/2}^{I=1}$ system in Fig.~\ref{Potential2}($b$),
the $\mathcal{O}(\varepsilon^0)$ contact interaction vanishes in our
calculation, and the total potential arises from the
two-pion-exchange contribution. We notice the potential in this
channel is much shallower than that of the $[DN]_{J=1/2}^{I=0}$
channel, i.e., the attraction is too feeble to form the bound state.
Thus the binding solution does not exist in this channel.

{\it $D^\ast N$ system}: The contact potential of the $[D^\ast
N]_{J=1/2}^{I=0}$ system in Fig.~\ref{Potential2}($c$) is
attractive, while the one-pion-exchange and two-pion-exchange
interaction are both repulsive. Therefore, the total potential is
shallower than that of the $[DN]_{J=1/2}^{I=0}$ channel. However, we
still obtain a binding solution in the $[D^\ast N]_{J=1/2}^{I=0}$
system. The binding energy and mass of this state are
\begin{eqnarray}
\Delta E_{[D^\ast N]_{J=1/2}^{I=0}}&\simeq&-1.5~\textrm{MeV},\nonumber\\
M_{[D^\ast N]_{J=1/2}^{I=0}}&\simeq&2943.6~\textrm{MeV}.
\end{eqnarray}

For the $[D^\ast N]_{J=1/2}^{I=1}$ system in
Fig.~\ref{Potential2}($d$), the one-pion-exchange potential is
weakly attractive, but the contact and two-pion-exchange potentials
are all repulsive. Thus, the total attractive potential is not
strong enough to form molecular states in this channel.

For the channel $[D^\ast N]_{J=3/2}^{I=0}$ in
Fig.~\ref{Potential2}($e$), the behavior of its potentials is very
interesting. We notice the one-pion- and two-pion-exchange
contributions almost cancel each other. Thus the total potential is
mainly provided by the contact term, which can reach up to $-80$ MeV
at the deepest position. By solving the Schr\"odinger equation, we
find the binding solution in the $[D^\ast N]_{J=3/2}^{I=0}$ system,
and the binding energy is
\begin{eqnarray}
\Delta E_{[D^\ast N]_{J=3/2}^{I=0}}\simeq-6.7~\textrm{MeV}.
\end{eqnarray}
The corresponding mass of this bound state is
\begin{eqnarray}
M_{[D^\ast N]_{J=3/2}^{I=0}}\simeq 2938.4~\textrm{MeV},
\end{eqnarray}
which is in good agreement with the mass of $\Lambda_c(2940)$
measured by BaBar, Belle and LHCb (e.g., see
Table~\ref{LambdacInfo}).

For the last channel $[D^\ast N]_{J=3/2}^{I=1}$ in
Fig.~\ref{Potential2}($f$), the one-pion- and two-pion-exchange
potentials almost cancel each other, and the contact contribution is
very weakly attractive. Thus no bound state can be found in this
channel.

{\it Role of the $\Delta(1232)$}: Considering the strong coupling
between $\Delta(1232)$ and $N\pi$, we include the contribution of
$\Delta(1232)$ in the loop diagrams (e.g., see
Figs.~\ref{TwoPion_Loop1} and \ref{TwoPion_Loop2}). Here, we discuss
the role of $\Delta(1232)$ in the the effective potentials of the
$DN$ and $D^\ast N$ systems. We try to ignore the effect of
$\Delta(1232)$, and notice that the lineshape of the
two-pion-exchange potentials changes drastically. Except for the
$[DN]_{J=1/2}^{I=1}$, the whole behavior of the other channels is
totally reversed. For example, the two-pion-exchange potential of
the $[DN]_{J=1/2}^{I=0}$ channel becomes repulsive, which renders
the total potential of this channel shallower. But for the $[D^\ast
N]_{J=3/2}^{I=0}$ channel, the two-pion-exchange potential becomes
attractive. The variation is about $-30$ MeV, which gives rise to a
deeper attractive potential, and the binding energy is $-16$ MeV.

In general, the conclusion that there exists the bound state in
isoscalar $[D^{(\ast)} N]_{J}$ channel and no binding solution in
isovector channel is robust, no matter we consider the
$\Delta(1232)$ or not. However, the $\Delta(1232)$ plays an
important role in determining the physical masses of
$\Lambda_c(2940)$ and other bound states, since the molecular states
are very sensitive to the subtle changes of their internal effective
potentials.

It is also interesting to see the dependence of the potentials on cutoff
$\Lambda$. The results for $\Lambda=0.6$ GeV are shown as the dashed lines
in Fig. \ref{Potential2}. We notice the change is dramatic, the total potentials are sensitive to the
cutoff, and they all become much deeper when the cutoff is varied to $0.6$ GeV. The binding energies for
$[D N]_{J=1/2}^{I=0}$, $[D^\ast N]_{J=1/2}^{I=0}$ and $[D^\ast N]_{J=3/2}^{I=0}$ channels in this case are
$-76.7$, $-11.6$ and $-55.1$ MeV, respectively. The isovector $DN$ system also starts to form bound state with binding enery
$-1.0$ MeV, but we still cannot get binding solutions in the isovector $D^\ast N$ system.
Of course, this behavior is foreseeable, because we introduced a Gaussian regulator when making the Fourier
transformation [see Eq. \eqref{Four_Tansf}]. Even in the $NN$ interactions, the LECs show mild dependence on the cutoff over a narrow range
only when the higher order contributions are included~\cite{Marji:2013uia,Entem:2017gor}. Generally, the LECs are used to absorb the scale dependence
of observables when making the nonperturbative renormalizations, i.e., they are usually the functions of cutoff $\Lambda$. Therefore, varying the cutoff
with the fixed LECs might be unreasonable. In the Appendix \ref{app:QM}, we illustrate the results with the LECs fixed at $\Lambda=0.6$ GeV from Ref.~\cite{Kang:2013uia}.

Inspecting the potentials in Fig.~\ref{Potential2} one can find two defects, one is the long-range behavior of the one-pion-exchange potential is missing, another one is
a bump appeared in the total potentials. Both are caused by the artificial effect from the Gaussian regulator. As indicated in Refs.~\cite{Epelbaum:2014efa,Reinert:2017usi},
such a form regulator would distort the partial waves and affect the long-range parts of the interactions. If the cutoff is sufficiently large, the induced artefacts are
expected to go beyond the accuracy at the order we are conducting. However, they may become an issue when the cutoff resides in the low scale. Additionally,
the $\mathcal{F}(\bm q)$ in Eq.~\eqref{Four_Tansf} has the risk of mixing different partial waves. But for the $S$-wave, the mixing effect is insufficient to impact the main feature
of the potentials.
\subsection{Discussions}

No binding solution in the isovector channel indicates that the
$[DN]_{J=1/2}^{I=1}$ molecular explanation of $\Sigma_c(2800)$ is
not favored. Although the $\Sigma_c(2800)$ is near the $DN$
threshold, its mass is also consistent with the quark model
predictions~\cite{Ebert:2011kk,Chen:2016iyi,Garcilazo:2007eh,Yoshida:2015tia,Ebert:2007nw}.
Thus interpreting the $\Sigma_c(2800)$ as the $1P$ charmed baryon
seems to be more reasonable.

The situation of $\Lambda_c(2940)$ is very similar to the
$\Lambda(1405)$, $D_{s0}(2317)$ and $X(3872)$, i.e., there is large
gap between the physical states and quark model
predictions\footnote{An unquenched study with the channel coupling
in Ref.~\cite{Luo:2019qkm} declares the mass of
$\Lambda_c(2P,3/2^-)$ state can be lowered down to match the
experimental data of $\Lambda_c(2940)$.}. Generally, one possible
reason is these states {\it per se} may be exotic rather than the
conventional ones.

A recent analysis from LHCb gives weak constraints on the $J^P$
quantum numbers of $\Lambda_c(2940)$, where $J^P=\frac{3}{2}^-$ is
favored~\cite{Aaij:2017vbw}. This is consistent with our
calculations. Actually, one can notice two peaks in the $D^0p$
invariant mass spectrum from $2.92$ GeV to $2.99$ GeV in the results
of LHCb (see Fig.~13(a) in Ref.~\cite{Aaij:2017vbw}). The one at
$2.94$ GeV is just the reported $\Lambda_c(2940)$. The other peak at
$2.98$ GeV may correspond to the true $\Lambda_c(2P)$ baryon, since
its mass is close to the quark model
prediction~\cite{Capstick:1986bm,Ebert:2011kk,Chen:2014nyo,Lu:2016ctt}.
Our calculation indicates the $\Lambda_c(2940)$ is probably the
$S$-wave $D^\ast N$ molecular state.

We report three bound states in the $[DN]_{J=1/2}^{I=0}$, $[D^\ast
N]_{J=1/2}^{I=0}$ and $[D^\ast N]_{J=3/2}^{I=0}$ systems. They are
very similar to the newly observed $P_c(4312)$, $P_c(4440)$ and
$P_c(4457)$ at LHCb~\cite{Aaij:2019vzc}, which are interpreted as
the $[\bar{D}\Sigma_c]_{J=1/2}^{I=1/2}$,
$[\bar{D}^\ast\Sigma_c]_{J=1/2}^{I=1/2}$ and
$[\bar{D}^\ast\Sigma_c]_{J=3/2}^{I=1/2}$ molecular
states~\cite{Meng:2019ilv,Wang:2019ato}, respectively. If
$\Lambda_c(2940)$ is indeed the $D^\ast N$ molecular states, then it
should contain two structures, i.e., $[D^\ast N]_{J=1/2}^{I=0}$ and
$[D^\ast N]_{J=3/2}^{I=0}$. Because the mass splitting between the
spin-$\frac{1}{2}$ and spin-$\frac{3}{2}$ states is only about $5$
MeV, it is very difficult to disassemble these two structures with
current accuracy. Similar situation has happened to the $P_c$
states. The previously reported $P_c(4450)$~\cite{Aaij:2015tga}
contains two structures, $P_c(4440)$ and $P_c(4457)$, after
increasing the data sample. More interesting, we find the mass of
the spin-$\frac{1}{2}$ state is larger than that of the
spin-$\frac{3}{2}$ one.

The signal of $\Lambda_c(2940)$ has been observed in the $D^0p$ and
$\Sigma_c\pi$ final
states~\cite{Aubert:2006sp,Abe:2006rz,Aaij:2017vbw}. However, if the
$J^P$ of $\Lambda_c(2940)$ is $\frac{3}{2}^-$ as weakly constrained
by the LHCb, then it decays into the $D^0p$ and $\Sigma_c\pi$
through the $D$-wave, which is strongly suppressed\footnote{Based on
the $^3P_0$ model calculation, the large decay width of
$\Lambda_c(2940)\to D^0p$ is reported in
Refs.~\cite{Lu:2018utx,Lu:2019rtg} by treating $\Lambda_c(2940)$ as
a $2P$ state in the $\Lambda_c$ family, but the low mass puzzle
still exist.}. Therefore, as mentioned above, one promising
explanation is that the $\Lambda_c(2940)$ signal actually contains
two structures. The spin-$\frac{1}{2}$ structure can easily decay
into $D^0p$ and $\Sigma_c\pi$ via the $S$-wave.

Borrowing experiences from the discovery of $P_c$ states, we urge
the experimenters to reanalyze the $\Lambda_c^+\pi^+\pi^-$ invariant
mass spectrum with the accumulated data, since the
$[DN]_{J=1/2}^{I=0}$, $[D^\ast N]_{J=1/2}^{I=0}$ and $[D^\ast
N]_{J=3/2}^{I=0}$ bound states can all decay into $\Sigma_c^0\pi^+$.

In addition to the mass spectrum, the decay pattern can also give us
some important criteria to identify the inner structure of
$\Lambda_c(2940)$. In the molecular scenario, the $D^\ast N$ system
can easily decay into the $DN$ channel via the pion exchange, while
the $\Sigma_c\pi$ decay mode requires the exchange of a nucleon or a
$D$ meson. Thus, the decay amplitude of the $D^0p$ mode should be
much larger than that of the $\Sigma_c\pi$, because the heavy hadron
exchange is generally suppressed. However, the phase space of the
$\Sigma_c\pi$ mode is larger.

The three body decay mode is also very interesting. We take the
decay modes of the $X(3872)$ and other higher charmonia as an
example. The branching fraction of $X(3872)\to D^0\bar{D}^0\pi^0$
can reach up to $40\%$~\cite{Tanabashi:2018oca}. In contrast, the
open charm three body decays of the higher charmonia is only a few
percents ~\cite{Weng:2018ebv}. Analogously, the branching fraction
of $\Lambda_c(2940)\to D^0\pi^0(\gamma) p$ should also be
conspicuous in the molecular picture.

Besides, our study can be easily extended to the $\bar{B}^{(\ast)}N$
systems. The axial coupling $g$ and mass splitting $\delta_b$ in
Eq.~\eqref{Meson_Lag_SF} should be replaced by the bottomed ones,
where we adopt $g=-0.52$~\cite{Ohki:2008py,Detmold:2012ge} and
$\delta_b=45$ MeV~\cite{Tanabashi:2018oca}. The predicted results
are listed in Table.~\ref{PredRes}. There also exist bound states in
the isoscalar $[\bar{B}^{(\ast)}N]_{J}$ systems. These states might
be reconstructed at the $\Lambda_b^0\pi^+\pi^-$ final states, and
the $[\bar{B}^{\ast}N]_{J}$ states could also be detected in the
$B^-p$ mass spectrum.
\begin{table}[htbp]
\centering
\renewcommand{\arraystretch}{1.8}
\caption{The predicted binding energies and masses for the isoscalar
$[D^{(\ast)}N]_J$ and $[\bar{B}^{(\ast)}N]_J$ systems (in units of
MeV).\label{PredRes}} \setlength{\tabcolsep}{0.9mm} {
\begin{tabular}{c|ccc|ccc}
\hline\hline
System&$[DN]_{\frac{1}{2}}$&$[D^\ast N]_{\frac{1}{2}}$&$[D^\ast N]_{\frac{3}{2}}$&$[\bar{B}N]_{\frac{1}{2}}$&$[\bar{B}^\ast N]_{\frac{1}{2}}$&$[\bar{B}^\ast N]_{\frac{3}{2}}$\\
\hline
$\Delta E$&$-11.1$&$-1.5$&$-6.7$&$-8.8$&$-3.5$&$-8.4$\\
Mass&$2792.0$&$2943.6$&$2938.4$&$6208.8$&$6259.4$&$6254.5$\\
\hline\hline
\end{tabular}
}
\end{table}

\section{Summary}\label{Summary}
A sophisticated investigation on the $DN$ and $D^\ast N$
interactions is crucial to clarify the nature of the charmed baryons
$\Sigma_c(2800)$ and $\Lambda_c(2940)$. In this work, we
systematically study the effective potentials of the $DN$ and
$D^\ast N$ systems with the chiral effective field theory up to the
next-to-leading order. We simultaneously consider the contributions
of the long-range one-pion-exchange, mid-range two-pion-exchange and
short-range contact term. We also include the $\Delta(1232)$ as an
intermediate state in the loop diagrams. The LECs are estimated from
the $N\bar{N}$ interaction with the help of quark model.

For the $DN$ system, our calculation shows the effective potentials
of the $[DN]_{J=1/2}^{I=0}$ and $[DN]_{J=1/2}^{I=1}$ channels are
both attractive. We find a bound state in the $[DN]_{J=1/2}^{I=0}$
channel, but the attraction in the $[DN]_{J=1/2}^{I=1}$ channel is
too weak to form a bound state. Thus the explanation of
$\Sigma_c(2800)$ as the $DN$ molecular state is disfavored in our
calculations. The $\Sigma_c(2800)$ is more likely to be the
conventional $1P$ charmed baryon, since its mass is well consistent
with the quark model prediction.

There are four channels in the $[D^\ast N]_{J}^{I}$ system. We find
only the isoscalar $[D^\ast N]_{J}$ potential is deep enough to form
the molecular state. We obtain the masses of the bound states in the
$[D^\ast N]_{J=1/2}^{I=0}$ and $[D^\ast N]_{J=3/2}^{I=0}$ channels
to be $2943.6$ and $2938.4$ MeV, respectively, which well accord
with the BaBar, Belle and LHCb measurements for $\Lambda_c(2940)$.
Considering the small mass splitting between the spin-$\frac{1}{2}$
and spin-$\frac{3}{2}$ states, we conjecture the $\Lambda_c(2940)$
signal contains two structures.

It is not so easy to squeeze the $\Lambda_c(2940)$ into the
conventional charmed baryon spectrum, since the $60$$-$$100$ MeV gap
between the physical mass and quark model prediction cannot be
readily remedied. However, this problem can be easily reconciled in
the molecular picture, i.e., the $\Lambda_c(2940)$ is probably the
isoscalar $D^\ast N$ molecule rather than the $2P$ charmed baryon.

We also investigate the influence of $\Delta(1232)$ in the loop
diagrams. The binding solutions always exist in the isoscalar
$[D^\ast N]_{J}$ channels no matter we include the $\Delta(1232)$ or
not. There still do not exist bound states in the isovector channels
even we ignore the $\Delta(1232)$. However, the $\Delta(1232)$ is
important in yielding the shallowly bound isoscalar $[D^{(\ast)}
N]_{J}$ states.

We hope experimentalist could seek for the pentaquark candidates in
the open charmed channels, where the $D^{(\ast)} N$ molecular
pentaquarks in the isoscalar systems might be reconstructed at the
$\Lambda_c^+\pi^+\pi^-$ final state.

\section*{Acknowledgments}
B. Wang is very grateful to X. Z. Weng for discussions on the
charmed baryon spectroscopy. This project is supported by the
National Natural Science Foundation of China under Grant 11975033.

\begin{appendix}

\section{Determining the LECs from $N\bar{N}$ interaction}\label{app:QM}

One needs to know the values of the LECs in Eq.~\eqref{Lag_NH} to
study the strength of the short-range interaction. As proposed in
Refs.~\cite{Meng:2019nzy,Wang:2019nvm} (more details can be found in
the appendix of these two references), the LECs of $D^{(\ast)}N$
systems can be bridged to those of the $N\bar{N}$ interaction with
the help of quark model. The way is analogous to the resonance
saturation model~\cite{Epelbaum:2001fm}, but we build the quark
level Lagrangian. We assume the contact interaction stems from the
heavy meson exchanging. We introduce $\mathscr{S}$ and
$\mathscr{A}^\mu$ to produce the central potential and spin-spin
interaction, respectively. The matrix form of $\mathscr{S}$ and
$\mathscr{A}^\mu$ can be expressed as
\begin{eqnarray}
\mathscr{S}&=&\mathscr{S}_3^i\tau^i+\sqrt{\frac{1}{3}}\mathscr{S}_1,\\
\mathscr{A}^\mu&=&\mathscr{A}_3^{\mu
i}\tau^i+\sqrt{\frac{1}{3}}\mathscr{A}_1^\mu,
\end{eqnarray}
where $\mathscr{S}_3$ ($\mathscr{A}_3^{\mu}$) and $\mathscr{S}_1$
($\mathscr{A}_1$) denote the isospin triplet and isospin singlet,
respectively. The coefficient $\sqrt{1\over 3}$ is introduced to
satisfy the SU(3) flavor symmetry.

The $q\bar{q}$ contact potential can be written as
\begin{eqnarray}\label{qqquark}
V_{q\bar{q}}&=&c_s(1-3\bm{\tau}_1\cdot\bm{\tau}_2)+c_t(1-3\bm{\tau}_1\cdot\bm{\tau}_2)\bm{\sigma}_1\cdot\bm{\sigma}_2,
\end{eqnarray}
where $c_s$ and $c_t$ are the coupling constants. The minus sign in
Eq.~\eqref{qqquark} arises since the isospin triplet and the isospin
singlet have the different $G$-parities.

With the $q\bar{q}$ contact potential $V_{q\bar{q}}$ in
Eq.~\eqref{qqquark} and the relevant matrix element in
Table~\ref{matrixelement}, we obtain the $N\bar{N}$ contact
potential as follows,
\begin{eqnarray}\label{NNbquark}
V_{N\bar{N}}&=&\langle N\bar{N}|V_{q\bar{q}}|N\bar{N}\rangle=9c_s-3c_s\bm{\tau}_1\cdot\bm{\tau}_2\nonumber\\
&&+c_t\bm{\sigma}_1\cdot\bm{\sigma}_2-\frac{25}{3}c_t(\bm{\tau}_1\cdot\bm{\tau}_2)(\bm{\sigma}_1\cdot\bm{\sigma}_2).
\end{eqnarray}

Similarly, the $D^\ast N$ contact potential can be easily worked
out,
\begin{eqnarray}\label{DastNquark}
V_{D^\ast N}&=&\langle D^\ast N|V_{q\bar{q}}|D^\ast N\rangle=3c_s-12c_s\mathbf{I}_1\cdot\mathbf{I}_2\nonumber\\
&&-c_t\bm{\sigma}\cdot\mathbf{T}+20c_t(\mathbf{I}_1\cdot\mathbf{I}_2)(\bm{\sigma}\cdot\mathbf{T}).
\end{eqnarray}

Matching Eq.~\eqref{Ct_DastN} and Eq.~\eqref{DastNquark} one can get
the LECs in Eq.~\eqref{Lag_NH}, which read
\begin{eqnarray}
D_a=3c_s,~~ D_b=-c_t,~~ E_a=3c_s,~~ E_b=-5c_t.
\end{eqnarray}

Therefore, once we know the values of $c_s$ and $c_t$, we can
capture the short range interaction of the $D^{(\ast)} N$ systems.
The $c_s$ and $c_t$ can be extracted from the $N\bar{N}$
interaction, and the $N\bar{N}$ scattering phase shift has been
fitted in the framework of chiral effective field theory to the
next-to-next-to-leading order in Ref.~\cite{Kang:2013uia}. Using the
values of $\tilde{C}_{^3S_1}$ in the $I=0$ and $I=1$ channels fitted
at $(\Lambda,\tilde{\Lambda})=(450,500)$ MeV as inputs, we obtain
\begin{eqnarray}\label{a7}
c_s=-8.1~\mathrm{GeV}^{-2},\qquad c_t=0.65~\mathrm{GeV}^{-2}.
\end{eqnarray}
We notice $|c_s|/|c_t|\simeq12.5$, i.e., the spin-spin interaction
only serves as a perturbation to give mass splittings between spin
multiplets.

\begin{table}[htbp]
\centering
\renewcommand{\arraystretch}{1.8}
\caption{The matrix elements of the operator $\sum_{i\in h_a,j\in
h_b}\mathscr{O}_{ij}$, where $h_a$ and $h_b$ are two hadrons.
$\mathscr{O}_{ij}$ is the two-body interaction operator between
quarks.\label{matrixelement}} \setlength{\tabcolsep}{2.5mm} {
\begin{tabular}{ccccc}
\hline\hline
$\mathscr{O}_{ij}$&$\bm{1}_{ij}$&$\bm{\tau}_i\cdot\bm{\tau}_j$&$\bm{\sigma}_i\cdot\bm{\sigma}_j$&$(\bm{\tau}_i\cdot\bm{\tau}_j)(\bm{\sigma}_i\cdot\bm{\sigma}_j)$\\
\hline
$[N\bar{N}]^{I=1}_{J=0}$&$9$&$1$&$-3$&$-\frac{25}{3}$\\
$[D^\ast N]^{I=1}_{J=3/2}$&$3$&$1$&$1$&$\frac{5}{3}$\\
\hline\hline
\end{tabular}
}
\end{table}

In addition, the Ref.~\cite{Kang:2013uia} also gives the fitted results with cutoff combination $(\Lambda,\tilde{\Lambda})=(650,700)$ MeV. With the
LECs of $\tilde{C}_{^3S_1}$ in the $I=0$ and $I=1$ channels fitted at $(\Lambda,\tilde{\Lambda})=(650,700)$ MeV as inputs we obtain
\begin{eqnarray}\label{a8}
c_s=-11.1~\mathrm{GeV}^{-2},\qquad c_t=4.5~\mathrm{GeV}^{-2}.
\end{eqnarray}

We see the $c_s$ value is similar to the one in Eq.~\eqref{a7}, while $c_t$ becomes much larger (We find that using other channel combinations in Ref.~\cite{Kang:2013uia} as inputs always obtains the $c_s$ value with the similar size and the same sign, but $c_t$ varies a lot). With the LECs in Eq.~\eqref{a8} as input we get the binding energies of the isoscalar and isovector $D^{(\ast)}N$ system as listed in Table~\ref{PredRes1}. One can notice that the binding energies of $[DN]_{J=1/2}^{I=0}$ and $[D^\ast N]_{J=1/2}^{I=0}$ channels are quite large, which can reach up to hundreds of MeV. This is highly unnatural in our framework, so we intend to use the moderate LECs in Eq.~\eqref{a7} rather than the ones in Eq.~\eqref{a8} to give predictions. However, we still can capture a gross indication that the isoscalar channel is always bound, while the isovector channel is not (except for the $[DN]_{J=1/2}^{I=1}$ channel).

\begin{table}[htbp]
\centering
\renewcommand{\arraystretch}{1.8}
\caption{The predicted binding energies for the isoscalar and isovector
$[D^{(\ast)}N]_J$ system (in units of
MeV) in the case of $c_s=-11.1$ GeV$^{-2}$, $c_t=4.5$ GeV$^{-2}$ and $\Lambda=0.6$ GeV. The ``$\times$" denotes no binding solution.\label{PredRes1}} \setlength{\tabcolsep}{0.9mm} {
\begin{tabular}{c|ccc|ccc}
\hline\hline
\multirow{2}{*}{System}&\multicolumn{3}{c|}{$I=0$}&\multicolumn{3}{c}{$I=1$}\\
&$[DN]_{\frac{1}{2}}$&$[D^\ast N]_{\frac{1}{2}}$&$[D^\ast N]_{\frac{3}{2}}$&$[DN]_{\frac{1}{2}}$&$[D^\ast N]_{\frac{1}{2}}$&$[D^\ast N]_{\frac{3}{2}}$\\
\hline
$\Delta E$&$-142.3$&$-294.8$&$-19.2$&$-1.0$&$\times$&$\times$\\
\hline\hline
\end{tabular}
}
\end{table}

\end{appendix}


\begin{thebibliography}{199}

\bibitem{Tanabashi:2018oca}
  M.~Tanabashi {\it et al.} [Particle Data Group],
  Review of Particle Physics,
  \href{https://journals.aps.org/prd/abstract/10.1103/PhysRevD.98.030001}{Phys.\ Rev.\ D {\bf 98}, 030001 (2018)}.


\bibitem{Choi:2003ue}
  S.~K.~Choi {\it et al.} [Belle Collaboration],
  Observation of a narrow charmonium-like state in exclusive $B^\pm\to K^\pm\pi^+\pi^-J/\psi$ decays,
  \href{https://journals.aps.org/prl/abstract/10.1103/PhysRevLett.91.262001}{Phys.\ Rev.\ Lett.\  {\bf 91}, 262001 (2003)}.

\bibitem{Aubert:2003fg}
  B.~Aubert {\it et al.} [BaBar Collaboration],
  Observation of a narrow meson decaying to $D_s^+ \pi^0$ at a mass of $2.32$-GeV/c$^2$,
  \href{https://journals.aps.org/prl/abstract/10.1103/PhysRevLett.90.242001}{Phys.\ Rev.\ Lett.\  {\bf 90}, 242001 (2003)}.

\bibitem{Chen:2016qju}
  H.~X.~Chen, W.~Chen, X.~Liu and S.~L.~Zhu,
  The hidden-charm pentaquark and tetraquark states,
  \href{http://www.sciencedirect.com/science/article/pii/S037015731630103X?via%3Dihub}{Phys.\ Rept.\  {\bf 639}, 1 (2016)}.


\bibitem{Guo:2017jvc}
  F.~K.~Guo, C.~Hanhart, U.~G.~Mei\ss ner, Q.~Wang, Q.~Zhao and B.~S.~Zou,
  Hadronic molecules,
  \href{https://journals.aps.org/rmp/abstract/10.1103/RevModPhys.90.015004}{Rev.\ Mod.\ Phys.\  {\bf 90}, 015004 (2018)}.

\bibitem{Liu:2019zoy}
  Y.~R.~Liu, H.~X.~Chen, W.~Chen, X.~Liu and S.~L.~Zhu,
  Pentaquark and tetraquark states,
  \href{https://www.sciencedirect.com/science/article/pii/S0146641019300304?via%3Dihub}{Prog.\ Part.\ Nucl.\ Phys.\  {\bf 107}, 237 (2019)}.

\bibitem{Lebed:2016hpi}
  R.~F.~Lebed, R.~E.~Mitchell and E.~S.~Swanson,
  Heavy-quark QCD exotica,
  \href{https://www.sciencedirect.com/science/article/pii/S0146641016300734?via%3Dihub}{Prog.\ Part.\ Nucl.\ Phys.\  {\bf 93}, 143 (2017)}.

\bibitem{Esposito:2016noz}
  A.~Esposito, A.~Pilloni and A.~D.~Polosa,
  Multiquark resonances,
  \href{https://www.sciencedirect.com/science/article/pii/S037015731630391X?via%3Dihub}{Phys.\ Rept.\  {\bf 668}, 1 (2017)}.

\bibitem{Brambilla:2019esw}
  N.~Brambilla, S.~Eidelman, C.~Hanhart, A.~Nefediev, C.~P.~Shen, C.~E.~Thomas, A.~Vairo and C.~Z.~Yuan,
  The $XYZ$ states: experimental and theoretical status and perspectives,
  \href{https://arxiv.org/abs/1907.07583}{arXiv:1907.07583}.

\bibitem{Aubert:2006sp}
  B.~Aubert {\it et al.} [BaBar Collaboration],
  Observation of a charmed baryon decaying to $D^0p$ at a mass near $2.94$ GeV/c$^2$,
  \href{https://journals.aps.org/prl/abstract/10.1103/PhysRevLett.98.012001}{Phys.\ Rev.\ Lett.\  {\bf 98}, 012001 (2007)}.

\bibitem{Abe:2006rz}
  K.~Abe {\it et al.} [Belle Collaboration],
  Experimental constraints on the spin and parity of the $\Lambda_c^+(2880)$,
  \href{https://journals.aps.org/prl/abstract/10.1103/PhysRevLett.98.262001}{Phys.\ Rev.\ Lett.\  {\bf 98}, 262001 (2007)}.

\bibitem{Aaij:2017vbw}
  R.~Aaij {\it et al.} [LHCb Collaboration],
  Study of the $D^0 p$ amplitude in $\Lambda_b^0\to D^0 p \pi^-$ decays,
  \href{https://link.springer.com/article/10.1007%2FJHEP05%282017%29030}{JHEP {\bf 1705}, 030 (2017)}.

\bibitem{Capstick:1986bm}
  S.~Capstick and N.~Isgur,
  Baryons in a relativized quark model with chromodynamics,
  \href{https://journals.aps.org/prd/abstract/10.1103/PhysRevD.34.2809}{Phys.\ Rev.\ D {\bf 34}, 2809 (1986)}.

\bibitem{Ebert:2011kk}
  D.~Ebert, R.~N.~Faustov and V.~O.~Galkin,
  Spectroscopy and Regge trajectories of heavy baryons in the relativistic quark-diquark picture,
  \href{https://journals.aps.org/prd/abstract/10.1103/PhysRevD.84.014025}{Phys.\ Rev.\ D {\bf 84}, 014025 (2011)}.

\bibitem{Chen:2014nyo}
  B.~Chen, K.~W.~Wei and A.~Zhang,
  Assignments of $\Lambda_Q$ and $\Xi_Q$ baryons in the heavy quark-light diquark picture,
  \href{https://link.springer.com/article/10.1140%2Fepja%2Fi2015-15082-3}{Eur.\ Phys.\ J.\ A {\bf 51}, 82 (2015)}.

\bibitem{Lu:2016ctt}
  Q.~F.~Lü, Y.~Dong, X.~Liu and T.~Matsuki,
  Puzzle of the $\Lambda_c$ spectrum,
  \href{http://www.npr.ac.cn/CN/10.11804/NuclPhysRev.35.01.001}{Nucl.\ Phys.\ Rev.\  {\bf 35}, 1 (2018)}.


\bibitem{He:2006is}
  X.~G.~He, X.~Q.~Li, X.~Liu and X.~Q.~Zeng,
  $\Lambda_c(2940)^+$: A Possible molecular state?,
  \href{https://link.springer.com/article/10.1140%2Fepjc%2Fs10052-007-0347-y}{Eur.\ Phys.\ J.\ C {\bf 51}, 883 (2007)}.

\bibitem{He:2010zq}
  J.~He, Y.~T.~Ye, Z.~F.~Sun and X.~Liu,
  The observed charmed hadron $\Lambda_c(2940)^+$ and the $D^*N$ interaction,
  \href{https://journals.aps.org/prd/abstract/10.1103/PhysRevD.82.114029}{Phys.\ Rev.\ D {\bf 82}, 114029 (2010)}.

\bibitem{Ortega:2012cx}
  P.~G.~Ortega, D.~R.~Entem and F.~Fernandez,
  Quark model description of the $\Lambda_c(2940)^+$ as a molecular $D^*N$ state and the possible existence of the $\Lambda_b(6248)$,
  \href{https://www.sciencedirect.com/science/article/pii/S0370269312012786?via%3Dihub}{Phys.\ Lett.\ B {\bf 718}, 1381 (2013)}.

\bibitem{Dong:2009tg}
  Y.~Dong, A.~Faessler, T.~Gutsche and V.~E.~Lyubovitskij,
  Strong two-body decays of the $\Lambda_c(2940)^+$ in a hadronic molecule picture,
  \href{https://journals.aps.org/prd/abstract/10.1103/PhysRevD.81.014006}{Phys.\ Rev.\ D {\bf 81}, 014006 (2010)}.

\bibitem{Dong:2010xv}
  Y.~Dong, A.~Faessler, T.~Gutsche, S.~Kumano and V.~E.~Lyubovitskij,
  Radiative decay of $\Lambda_c(2940)^+$ in a hadronic molecule picture,
  \href{https://journals.aps.org/prd/abstract/10.1103/PhysRevD.82.034035}{Phys.\ Rev.\ D {\bf 82}, 034035 (2010)}.

\bibitem{Zhang:2012jk}
  J.~R.~Zhang,
  $S$-wave $D^{(*)}N$ molecular states: $\Sigma_{c}(2800)$ and $\Lambda_{c}(2940)^{+}$?,
  \href{https://journals.aps.org/prd/abstract/10.1103/PhysRevD.89.096006}{Phys.\ Rev.\ D {\bf 89}, 096006 (2014)}.

\bibitem{Zhao:2016zhf}
  L.~Zhao, H.~Huang and J.~Ping,
  $ND$ and $NB$ systems in quark delocalization color screening model,
  \href{https://link.springer.com/article/10.1140%2Fepja%2Fi2017-12219-4}{Eur.\ Phys.\ J.\ A {\bf 53}, 28 (2017)}.

\bibitem{Zhang:2019vqe}
  D.~Zhang, D.~Yang, X.~F.~Wang and K.~Nakayama,
  Possible $S$-wave $ND^{(*)}$ and $N\bar B^{(*)}$ bound states in a chiral quark model,
  \href{https://arxiv.org/abs/1903.01207}{arXiv:1903.01207}.

\bibitem{Klempt:2009pi}
  E.~Klempt and J.~M.~Richard,
  Baryon spectroscopy,
  \href{https://journals.aps.org/rmp/abstract/10.1103/RevModPhys.82.1095}{Rev.\ Mod.\ Phys.\  {\bf 82}, 1095 (2010)}.

\bibitem{Crede:2013sze}
  V.~Crede and W.~Roberts,
  Progress towards understanding baryon resonances,
  \href{https://iopscience.iop.org/article/10.1088/0034-4885/76/7/076301}{Rept.\ Prog.\ Phys.\  {\bf 76}, 076301 (2013)}.

\bibitem{Cheng:2015iom}
  H.~Y.~Cheng,
  Charmed baryons circa 2015,
  \href{https://link.springer.com/article/10.1007%2Fs11467-015-0483-z}{Front.\ Phys.\ (Beijing) {\bf 10}, 101406 (2015)}.

\bibitem{Chen:2016spr}
  H.~X.~Chen, W.~Chen, X.~Liu, Y.~R.~Liu and S.~L.~Zhu,
  A review of the open charm and open bottom systems,
  \href{https://iopscience.iop.org/article/10.1088/1361-6633/aa6420}{Rept.\ Prog.\ Phys.\  {\bf 80}, 076201 (2017)}.

\bibitem{Kato:2018ijx}
  Y.~Kato and T.~Iijima,
  Open charm hadron spectroscopy at B-factories,
  \href{https://www.sciencedirect.com/science/article/pii/S0146641019300018?via%3Dihub}{Prog.\ Part.\ Nucl.\ Phys.\  {\bf 105}, 61 (2019)}.

  \bibitem{Huang:2016ygf}
  Y.~Huang, J.~He, J.~J.~Xie and L.~S.~Geng,
  Production of the $\Lambda_c(2940)$ by kaon-induced reactions on a proton target,
  \href{https://journals.aps.org/prd/abstract/10.1103/PhysRevD.99.014045}{Phys.\ Rev.\ D {\bf 99}, 014045 (2019)}.

\bibitem{Wang:2015rda}
  X.~Y.~Wang, A.~Guskov and X.~R.~Chen,
  $\Lambda_c^*(2940)^+$ photoproduction off the neutron,
  \href{https://journals.aps.org/prd/abstract/10.1103/PhysRevD.92.094032}{Phys.\ Rev.\ D {\bf 92}, 094032 (2015)}.

\bibitem{Cheng:2015naa}
  H.~Y.~Cheng and C.~K.~Chua,
  Strong decays of charmed baryons in heavy hadron chiral perturbation theory: An update,
  \href{https://journals.aps.org/prd/abstract/10.1103/PhysRevD.92.074014}{Phys.\ Rev.\ D {\bf 92}, 074014 (2015)}.

\bibitem{Xie:2015zga}
  J.~J.~Xie, Y.~B.~Dong and X.~Cao,
  Role of the $\Lambda^+_c(2940)$ in the $\pi^- p \to D^- D^0 p$ reaction close to threshold,
  \href{https://journals.aps.org/prd/abstract/10.1103/PhysRevD.92.034029}{Phys.\ Rev.\ D {\bf 92}, 034029 (2015)}.

\bibitem{Romanets:2012hm}
  O.~Romanets, L.~Tolos, C.~Garcia-Recio, J.~Nieves, L.~L.~Salcedo and R.~G.~E.~Timmermans,
  Charmed and strange baryon resonances with heavy-quark spin symmetry,
  \href{https://journals.aps.org/prd/abstract/10.1103/PhysRevD.85.114032}{Phys.\ Rev.\ D {\bf 85}, 114032 (2012)}.

\bibitem{He:2011jp}
  J.~He, Z.~Ouyang, X.~Liu and X.~Q.~Li,
  Production of charmed baryon $\Lambda_c(2940)^+$ at PANDA,
  \href{https://journals.aps.org/prd/abstract/10.1103/PhysRevD.84.114010}{Phys.\ Rev.\ D {\bf 84}, 114010 (2011)}.

\bibitem{Cheng:2006dk}
  H.~Y.~Cheng and C.~K.~Chua,
  Strong decays of charmed baryons in heavy hadron chiral perturbation theory,
  \href{https://journals.aps.org/prd/abstract/10.1103/PhysRevD.75.014006}{Phys.\ Rev.\ D {\bf 75}, 014006 (2007)}.

\bibitem{Chen:2007xf}
  C.~Chen, X.~L.~Chen, X.~Liu, W.~Z.~Deng and S.~L.~Zhu,
  Strong decays of charmed baryons,
  \href{https://journals.aps.org/prd/abstract/10.1103/PhysRevD.75.094017}{Phys.\ Rev.\ D {\bf 75}, 094017 (2007)}.

\bibitem{Zhong:2007gp}
  X.~H.~Zhong and Q.~Zhao,
  Charmed baryon strong decays in a chiral quark model,
  \href{https://journals.aps.org/prd/abstract/10.1103/PhysRevD.77.074008}{Phys.\ Rev.\ D {\bf 77}, 074008 (2008)}.

\bibitem{Chen:2015kpa}
  H.~X.~Chen, W.~Chen, Q.~Mao, A.~Hosaka, X.~Liu and S.~L.~Zhu,
  $P$-wave charmed baryons from QCD sum rules,
  \href{https://journals.aps.org/prd/abstract/10.1103/PhysRevD.91.054034}{Phys.\ Rev.\ D {\bf 91}, 054034 (2015)}.

\bibitem{Mizuk:2004yu}
  R.~Mizuk {\it et al.} [Belle Collaboration],
  Observation of an isotriplet of excited charmed baryons decaying to $\Lambda_c \pi$,
  \href{https://journals.aps.org/prl/abstract/10.1103/PhysRevLett.94.122002}{Phys.\ Rev.\ Lett.\  {\bf 94}, 122002 (2005)}.

\bibitem{Aubert:2008ax}
  B.~Aubert {\it et al.} [BaBar Collaboration],
  Measurements of $\mathscr{B}(\bar{B}^0\to\Lambda_c^+\bar{p})$ and $\mathscr{B}(\bar{B}^-\to\Lambda_c^+\bar{p}\pi^-)$ and studies of $\Lambda_c^+\pi^-$ resonances,
  \href{https://journals.aps.org/prd/abstract/10.1103/PhysRevD.78.112003}{Phys.\ Rev.\ D {\bf 78}, 112003 (2008)}.

\bibitem{Chen:2016iyi}
  B.~Chen, K.~W.~Wei, X.~Liu and T.~Matsuki,
  Low-lying charmed and charmed-strange baryon states,
  \href{https://link.springer.com/article/10.1140%2Fepjc%2Fs10052-017-4708-x}{Eur.\ Phys.\ J.\ C {\bf 77}, 154 (2017)}.

\bibitem{Garcilazo:2007eh}
  H.~Garcilazo, J.~Vijande and A.~Valcarce,
  Faddeev study of heavy baryon spectroscopy,
  \href{https://iopscience.iop.org/article/10.1088/0954-3899/34/5/014}{J.\ Phys.\ G {\bf 34}, 961 (2007)}.

\bibitem{Wang:2017kfr}
  K.~L.~Wang, Y.~X.~Yao, X.~H.~Zhong and Q.~Zhao,
  Strong and radiative decays of the low-lying $S$- and $P$-wave singly heavy baryons,
  \href{https://journals.aps.org/prd/abstract/10.1103/PhysRevD.96.116016}{Phys.\ Rev.\ D {\bf 96}, 116016 (2017)}.

\bibitem{Dong:2010gu}
  Y.~Dong, A.~Faessler, T.~Gutsche and V.~E.~Lyubovitskij,
  Charmed baryon $\Sigma_c(2800)$ as a $ND$ hadronic molecule,
  \href{https://journals.aps.org/prd/abstract/10.1103/PhysRevD.81.074011}{Phys.\ Rev.\ D {\bf 81}, 074011 (2010)}.

\bibitem{Tsushima:1998ru}
  K.~Tsushima, D.~H.~Lu, A.~W.~Thomas, K.~Saito and R.~H.~Landau,
  Charmed mesic nuclei,
  \href{https://journals.aps.org/prc/abstract/10.1103/PhysRevC.59.2824}{Phys.\ Rev.\ C {\bf 59}, 2824 (1999)}.

\bibitem{GarciaRecio:2010vt}
  C.~Garcia-Recio, J.~Nieves and L.~Tolos,
  $D$ mesic nuclei,
  \href{https://www.sciencedirect.com/science/article/pii/S0370269310006489?via%3Dihub}{Phys.\ Lett.\ B {\bf 690}, 369 (2010)}.

\bibitem{Hosaka:2016ypm}
  A.~Hosaka, T.~Hyodo, K.~Sudoh, Y.~Yamaguchi and S.~Yasui,
  Heavy hadrons in nuclear matter,
  \href{https://www.sciencedirect.com/science/article/pii/S0146641017300388?via%3Dihub}{Prog.\ Part.\ Nucl.\ Phys.\  {\bf 96}, 88 (2017)}.

\bibitem{Krein:2017usp}
  G.~Krein, A.~W.~Thomas and K.~Tsushima,
  Nuclear-bound quarkonia and heavy-flavor hadrons,
  \href{https://www.sciencedirect.com/science/article/pii/S0146641018300127?via%3Dihub}{Prog.\ Part.\ Nucl.\ Phys.\  {\bf 100}, 161 (2018)}.

\bibitem{Machleidt:1987hj}
  R.~Machleidt, K.~Holinde and C.~Elster,
  The Bonn meson exchange model for the nucleon-nucleon interaction,
  \href{https://www.sciencedirect.com/science/article/abs/pii/S0370157387800029?via%3Dihub}{Phys.\ Rept.\  {\bf 149}, 1 (1987)}.

\bibitem{Haidenbauer:2007jq}
  J.~Haidenbauer, G.~Krein, U.~G.~Mei\ss ner and A.~Sibirtsev,
  $\bar{D}N$ interaction from meson-exchange and quark-gluon dynamics,
  \href{https://link.springer.com/article/10.1140%2Fepja%2Fi2007-10417-3}{Eur.\ Phys.\ J.\ A {\bf 33}, 107 (2007)}.

\bibitem{Haidenbauer:2008ff}
  J.~Haidenbauer, G.~Krein, U.~G.~Mei\ss ner and A.~Sibirtsev,
  Charmed meson rescattering in the reaction $\bar{p}d\to \bar{D}DN$,
  \href{https://link.springer.com/article/10.1140%2Fepja%2Fi2008-10602-x}{Eur.\ Phys.\ J.\ A {\bf 37}, 55 (2008)}.

\bibitem{Haidenbauer:2010ch}
  J.~Haidenbauer, G.~Krein, U.~G.~Mei\ss ner and L.~Tolos,
  $DN$ interaction from meson exchange,
  \href{https://link.springer.com/article/10.1140%2Fepja%2Fi2011-11018-3}{Eur.\ Phys.\ J.\ A {\bf 47}, 18 (2011)}.

\bibitem{Weinberg:1990rz}
  S.~Weinberg,
  Nuclear forces from chiral Lagrangians,
  \href{http://www.sciencedirect.com/science/article/pii/0370269390909383?via%3Dihub}{Phys.\ Lett.\ B {\bf 251}, 288 (1990)}.

\bibitem{Weinberg:1991um}
  S.~Weinberg,
  Effective chiral Lagrangians for nucleon-pion interactions and nuclear forces,
  \href{http://www.sciencedirect.com/science/article/pii/055032139190231L?via%3Dihub}{Nucl.\ Phys.\ B {\bf 363}, 3 (1991)}.

\bibitem{Bernard:1995dp}
  V.~Bernard, N.~Kaiser and U.~G.~Mei\ss ner,
  Chiral dynamics in nucleons and nuclei,
  \href{http://www.worldscientific.com/doi/abs/10.1142/S0218301395000092}{Int.\ J.\ Mod.\ Phys.\ E {\bf 4}, 193 (1995)}.

\bibitem{Epelbaum:2008ga}
  E.~Epelbaum, H.~W.~Hammer and U.~G.~Mei\ss ner,
  Modern theory of nuclear forces,
  \href{https://journals.aps.org/rmp/abstract/10.1103/RevModPhys.81.1773}{Rev.\ Mod.\ Phys.\  {\bf 81}, 1773 (2009)}.

\bibitem{Machleidt:2011zz}
  R.~Machleidt and D.~R.~Entem,
  Chiral effective field theory and nuclear forces,
  \href{http://www.sciencedirect.com/science/article/pii/S0370157311000457?via%3Dihub}{Phys.\ Rept.\  {\bf 503}, 1 (2011)}.

\bibitem{Meissner:2015wva}
  U.~G.~Mei\ss ner,
  The long and winding road from chiral effective Lagrangians to nuclear structure,
  \href{http://iopscience.iop.org/article/10.1088/0031-8949/91/3/033005/meta}{Phys.\ Scripta {\bf 91}, 033005 (2016)}.


\bibitem{Hammer:2019poc}
  H.-W.~Hammer, S.~K\"onig and U.~van Kolck,
  Nuclear effective field theory: status and perspectives,
  \href{https://arxiv.org/abs/1906.12122}{arXiv:1906.12122}.

\bibitem{Machleidt:2020vzm}
  R.~Machleidt and F.~Sammarruca,
  Can chiral EFT give us satisfaction?,
  \href{https://arxiv.org/abs/2001.05615}{arXiv:2001.05615}.

\bibitem{Liu:2012vd}
  Z.~W.~Liu, N.~Li and S.~L.~Zhu,
  Chiral perturbation theory and the $\bar B \bar B$ strong interaction,
  \href{https://journals.aps.org/prd/abstract/10.1103/PhysRevD.89.074015}{Phys.\ Rev.\ D {\bf 89}, 074015 (2014)}.

\bibitem{Xu:2017tsr}
  H.~Xu, B.~Wang, Z.~W.~Liu and X.~Liu,
  $D D^{*}$ potentials in chiral perturbation theory and possible molecular states,
  \href{https://journals.aps.org/prd/abstract/10.1103/PhysRevD.99.014027}{Phys.\ Rev.\ D {\bf 99}, 014027 (2019)}.

\bibitem{Wang:2018atz}
  B.~Wang, Z.~W.~Liu and X.~Liu,
  $\bar{B}^{(\ast)} \bar{B}^{(\ast)}$ interactions in chiral effective field theory,
  \href{https://journals.aps.org/prd/abstract/10.1103/PhysRevD.99.036007}{Phys.\ Rev.\ D {\bf 99}, 036007 (2019)}.

\bibitem{Meng:2019ilv}
  L.~Meng, B.~Wang, G.~J.~Wang and S.~L.~Zhu,
  The hidden charm pentaquark states and $\Sigma_c\bar{D}^{(*)}$ interaction in chiral perturbation theory,
  \href{https://journals.aps.org/prd/abstract/10.1103/PhysRevD.100.014031}{Phys.\ Rev.\ D {\bf 100}, 014031 (2019)}.

\bibitem{Wang:2019ato}
  B.~Wang, L.~Meng and S.~L.~Zhu,
  Hidden-charm and hidden-bottom molecular pentaquarks in chiral perturbation theory,
  \href{https://link.springer.com/article/10.1007%2FJHEP11%282019%29108}{JHEP {\bf 1911}, 108 (2019)}.

\bibitem{Meng:2019nzy}
  L.~Meng, B.~Wang and S.~L.~Zhu,
  $\Sigma_cN$ interaction in chiral perturbation theory,
  \href{https://arxiv.org/abs/1912.09661}{arXiv:1912.09661}.

\bibitem{Wang:2019nvm}
  B.~Wang, L.~Meng and S.~L.~Zhu,
  Spectrum of the strange hidden charm molecular pentaquarks in chiral effective field theory,
  \href{https://journals.aps.org/prd/abstract/10.1103/PhysRevD.101.034018}{Phys.\ Rev.\ D {\bf 101}, 034018 (2020)}.

\bibitem{Scherer:2002tk}
  S.~Scherer,
  Introduction to chiral perturbation theory,
  \href{https://arxiv.org/abs/hep-ph/0210398}{Adv.\ Nucl.\ Phys.\  {\bf 27}, 277 (2003)}.


\bibitem{Holinde:1977rh}
  K.~Holinde and R.~Machleidt,
  Effect of the $\Delta(1236)$ resonance on $NN$ scattering, nuclear matter and neutron matter,
  \href{https://www.sciencedirect.com/science/article/abs/pii/0375947477906145?via%3Dihub}{Nucl.\ Phys.\ A {\bf 280}, 429 (1977)}.

\bibitem{Krebs:2007rh}
  H.~Krebs, E.~Epelbaum and U.~G.~Mei\ss ner,
  Nuclear forces with $\Delta$-excitations up to next-to-next-to-leading order, part I: Peripheral nucleon-nucleon waves,
  \href{https://link.springer.com/article/10.1140%2Fepja%2Fi2007-10372-y}{Eur.\ Phys.\ J.\ A {\bf 32}, 127 (2007)}.

\bibitem{Epelbaum:2007sq}
  E.~Epelbaum, H.~Krebs and U.~G.~Mei\ss ner,
  $\Delta$-excitations and the three-nucleon force,
  \href{https://www.sciencedirect.com/science/article/pii/S037594740800393X?via%3Dihub}{Nucl.\ Phys.\ A {\bf 806}, 65 (2008)}.

\bibitem{Kaiser:1998wa}
  N.~Kaiser, S.~Gerstendorfer and W.~Weise,
  Peripheral $NN$ scattering: Role of delta excitation, correlated two pion and vector meson exchange,
  \href{https://www.sciencedirect.com/science/article/pii/S0375947498002346?via%3Dihub}{Nucl.\ Phys.\ A {\bf 637}, 395 (1998)}.

\bibitem{Hemmert:1997ye}
  T.~R.~Hemmert, B.~R.~Holstein and J.~Kambor,
  Chiral Lagrangians and $\Delta(1232)$ interactions: Formalism,
  \href{https://iopscience.iop.org/article/10.1088/0954-3899/24/10/003}{J.\ Phys.\ G {\bf 24}, 1831 (1998)}.


\bibitem{Wise:1992hn}
  M.~B.~Wise,
  Chiral perturbation theory for hadrons containing a heavy quark,
  \href{https://journals.aps.org/prd/abstract/10.1103/PhysRevD.45.R2188}{Phys.\ Rev.\ D {\bf 45}, R2188 (1992)}.

\bibitem{Manohar:2000dt}
  A.~V.~Manohar and M.~B.~Wise,
  {\it Heavy quark physics},
  Camb.\ Monogr.\ Part.\ Phys.\ Nucl.\ Phys.\ Cosmol.\  {\bf 10}, 1 (2000).

\bibitem{Epelbaum:2004fk}
  E.~Epelbaum, W.~Glockle and U.~G.~Mei\ss ner,
  The two-nucleon system at next-to-next-to-next-to-leading order,
  \href{https://www.sciencedirect.com/science/article/pii/S0375947404010747?via%3Dihub}{Nucl.\ Phys.\ A {\bf 747}, 362 (2005)}.

\bibitem{Epelbaum:2003xx}
  E.~Epelbaum, W.~Gloeckle and U.~G.~Mei\ss ner,
  Improving the convergence of the chiral expansion for nuclear forces. 2. Low phases and the deuteron,
  \href{https://link.springer.com/article/10.1140%2Fepja%2Fi2003-10129-8}{Eur.\ Phys.\ J.\ A {\bf 19}, 401 (2004)}.

\bibitem{Kang:2013uia}
  X.~W.~Kang, J.~Haidenbauer and U.~G.~Mei\ss ner,
  Antinucleon-nucleon interaction in chiral effective field theory,
  \href{https://link.springer.com/article/10.1007%2FJHEP02%282014%29113}{JHEP {\bf 1402}, 113 (2014)}.

\bibitem{Epelbaum:2014efa}
  E.~Epelbaum, H.~Krebs and U.~G.~Mei\ss ner,
  Improved chiral nucleon-nucleon potential up to next-to-next-to-next-to-leading order,
  \href{https://link.springer.com/article/10.1140%2Fepja%2Fi2015-15053-8}{Eur.\ Phys.\ J.\ A {\bf 51}, 53 (2015)}.


\bibitem{Entem:2003ft}
  D.~R.~Entem and R.~Machleidt,
  Accurate charge dependent nucleon nucleon potential at fourth order of chiral perturbation theory,
  \href{https://journals.aps.org/prc/abstract/10.1103/PhysRevC.68.041001}{Phys.\ Rev.\ C {\bf 68}, 041001 (2003)}.

\bibitem{Dai:2017ont}
  L.~Y.~Dai, J.~Haidenbauer and U.~G.~Meißner,
  Antinucleon-nucleon interaction at next-to-next-to-next-to-leading order in chiral effective field theory,
  \href{https://link.springer.com/article/10.1007%2FJHEP07%282017%29078}{JHEP {\bf 1707}, 078 (2017)}.

\bibitem{Marji:2013uia}
  E.~Marji, A.~Canul, Q.~MacPherson, R.~Winzer, C.~Zeoli, D.~R.~Entem and R.~Machleidt,
  Nonperturbative renormalization of the chiral nucleon-nucleon interaction up to next-to-next-to-leading order,
  \href{https://journals.aps.org/prc/abstract/10.1103/PhysRevC.88.054002}{Phys.\ Rev.\ C {\bf 88}, 054002 (2013)}.

\bibitem{Entem:2017gor}
  D.~R.~Entem, R.~Machleidt and Y.~Nosyk,
  High-quality two-nucleon potentials up to fifth order of the chiral expansion,
  \href{https://journals.aps.org/prc/abstract/10.1103/PhysRevC.96.024004}{Phys.\ Rev.\ C {\bf 96}, 024004 (2017)}.


\bibitem{Reinert:2017usi}
  P.~Reinert, H.~Krebs and E.~Epelbaum,
  Semilocal momentum-space regularized chiral two-nucleon potentials up to fifth order,
  \href{https://link.springer.com/article/10.1140%2Fepja%2Fi2018-12516-4}{Eur.\ Phys.\ J.\ A {\bf 54}, 86 (2018)}.

\bibitem{Yoshida:2015tia}
  T.~Yoshida, E.~Hiyama, A.~Hosaka, M.~Oka and K.~Sadato,
  Spectrum of heavy baryons in the quark model,
  \href{https://journals.aps.org/prd/abstract/10.1103/PhysRevD.92.114029}{Phys.\ Rev.\ D {\bf 92}, 114029 (2015)}.

\bibitem{Ebert:2007nw}
  D.~Ebert, R.~N.~Faustov and V.~O.~Galkin,
  Masses of excited heavy baryons in the relativistic quark model,
  \href{https://www.sciencedirect.com/science/article/pii/S0370269307014402?via%3Dihub}{Phys.\ Lett.\ B {\bf 659}, 612 (2008)}.

\bibitem{Luo:2019qkm}
  S.~Q.~Luo, B.~Chen, Z.~W.~Liu and X.~Liu,
  Resolving the low mass puzzle of $\Lambda_c(2940)^+$,
  \href{https://link.springer.com/article/10.1140%2Fepjc%2Fs10052-020-7874-1}{Eur.\ Phys.\ J.\ C {\bf 80}, 301 (2020)}.


\bibitem{Aaij:2019vzc}
  R.~Aaij {\it et al.} [LHCb Collaboration],
  Observation of a narrow pentaquark state, $P_c(4312)^+$, and of two-peak structure of the $P_c(4450)^+$,
  \href{https://journals.aps.org/prl/abstract/10.1103/PhysRevLett.122.222001}{Phys.\ Rev.\ Lett.\  {\bf 122}, 222001 (2019)}.

\bibitem{Aaij:2015tga}
  R.~Aaij {\it et al.} [LHCb Collaboration],
  Observation of $J/\psi p$ resonances consistent with pentaquark states in $\Lambda_b^0 \to J/\psi K^- p$ decays,
  \href{https://journals.aps.org/prl/abstract/10.1103/PhysRevLett.115.072001}{Phys.\ Rev.\ Lett.\  {\bf 115}, 072001 (2015)}.

\bibitem{Lu:2018utx}
  Q.~F.~Lü, L.~Y.~Xiao, Z.~Y.~Wang and X.~H.~Zhong,
  Strong decay of $\Lambda _c(2940)$ as a $2P$ state in the $\Lambda _c$ family,
  \href{https://link.springer.com/article/10.1140%2Fepjc%2Fs10052-018-6083-7}{Eur.\ Phys.\ J.\ C {\bf 78}, 599 (2018)}.

\bibitem{Lu:2019rtg}
  Q.~F.~Lü and X.~H.~Zhong,
  Strong decays of the higher excited $\Lambda_Q$ and $\Sigma_Q$ baryons,
  \href{https://journals.aps.org/prd/abstract/10.1103/PhysRevD.101.014017}{Phys.\ Rev.\ D {\bf 101}, 014017 (2020)}.

\bibitem{Weng:2018ebv}
  X.~Z.~Weng, L.~Y.~Xiao, W.~Z.~Deng, X.~L.~Chen and S.~L.~Zhu,
  Three body open flavor decays of higher charmonium and bottomonium,
  \href{https://journals.aps.org/prd/abstract/10.1103/PhysRevD.99.094001}{Phys.\ Rev.\ D {\bf 99}, 094001 (2019)}.

\bibitem{Ohki:2008py}
  H.~Ohki, H.~Matsufuru and T.~Onogi,
  Determination of $B^\ast B \pi$ coupling in unquenched QCD,
  \href{https://journals.aps.org/prd/pdf/10.1103/PhysRevD.77.094509}{Phys.\ Rev.\ D {\bf 77}, 094509 (2008)}.

\bibitem{Detmold:2012ge}
  W.~Detmold, C.~J.~D.~Lin and S.~Meinel,
  Calculation of the heavy-hadron axial couplings $g_1$, $g_2$ and $g_3$ using lattice QCD,
  \href{https://journals.aps.org/prd/abstract/10.1103/PhysRevD.85.114508}{Phys.\ Rev.\ D {\bf 85}, 114508 (2012)}.


\bibitem{Epelbaum:2001fm}
  E.~Epelbaum, U.~G.~Mei\ss ner, W.~Gloeckle and C.~Elster,
  Resonance saturation for four nucleon operators,
  \href{https://journals.aps.org/prc/abstract/10.1103/PhysRevC.65.044001}{Phys.\ Rev.\ C {\bf 65}, 044001 (2002)}.

\end{thebibliography}
\end{document}